\documentclass[aps,prd,12point,twocolumn,nofootinbib,showpacs,superscriptaddress]{revtex4-2}

\usepackage{graphicx}% Include figure files
\usepackage{dcolumn}% Align table columns on decimal point
\usepackage{tabu}
\usepackage{amsmath,amssymb}
\usepackage{float}
\usepackage{natbib}
\usepackage{balance}
\usepackage[normalem]{ulem}
\usepackage{braket}
\usepackage{mathrsfs}
\usepackage{bm}% bold math
\usepackage{lipsum}
\usepackage{comment}
\usepackage{xfrac}
\usepackage[dvipsnames]{xcolor}
\usepackage[title]{appendix}
\usepackage{txfonts}
\usepackage{xcolor}
\usepackage{slashed}
\usepackage[breaklinks=true,pdfborder={0 0 0},bookmarksopen]{hyperref}
\hypersetup{
	colorlinks   = true, %Colours links instead of ugly boxes
	urlcolor     = magenta, %Colour for external hyperlinks
	linkcolor    = blue, %Colour of internal links
	citecolor   = Red %Colour of citations
}

\begin{document}

\title{Neutrino spin oscillation in screening models revisited}

\author{Fay\c{c}al Hammad}
\email{fhammad@ubishops.ca}
\affiliation{Department of Physics \& Astronomy,
Bishop's University,\\
2600 College Street, Sherbrooke, QC, J1M 1Z7, Canada}
\affiliation{Physics Department, Champlain College-Lennoxville,\\
2580 College Street, Sherbrooke, QC, J1M 0C8, Canada}

\author{Nicolas Fleury}
\email{nfleury22@ubishops.ca}
\affiliation{Department of Physics \& Astronomy,
Bishop's University,\\
2600 College Street, Sherbrooke, QC, J1M 1Z7, Canada}

\author{Parvaneh Sadeghi}
\email{psadeghi20@ubishops.ca}
\affiliation{Department of Physics \& Astronomy,
Bishop's University,\\
2600 College Street, Sherbrooke, QC, J1M 1Z7, Canada}
%%%%%%%%%%%%%%%%%%%%%%%%%%%%%%%%%%%%%%%%%%%%%%%%%%%%%%%%%%%%%%%%%%%%%%%%%%%%%%%%%%%%%%%%%%%%
%%%%%%%%%%%%%%%%%%%%%%%%%%%%%%%%%%%%%%%%%%%%%%%%%%%%%%%%%%%%%%%%%%%%%%%%%%%%%%%%%%%%%%%%%%%%
%%%%%%%%%%%%%%%%%%%%%%%%%%%%%%%%%%%%%%%%%%%%%%%%%%%%%%%%%%%%%%%%%%%%%%%%%%%%%%%%%%%%%%%%%%%%
%%%%%%%%%%%%%%%%%%%%%%%%%%%%%%%%%%%%%%%%%%%%%%%%%%%%%%%%%%%%%%%%%%%%%%%%%%%%%%%%%%%%%%%%%%%%

\begin{abstract}
We study, using the Mathisson-Papapetrou-Dixon equations, the spin oscillation of neutrinos when the latter are coupled to the scalar field of screening models of dark energy. First, we derive the transition probability formula for a left-handed neutrino to become a right-handed neutrino within a general static and spherically symmetric metric. We then apply our general formula to neutrinos deflected around a central mass described by the Schwarzschild metric. Our results show that, contrary to what one might expect, the scalar field of chameleon-like and symmetron-like screening models would not show any effect on the spin oscillations of neutrinos. The origin of such an outcome is discussed.
\end{abstract}

%\pacs {03.65.Pm, 04.62.+v, 04.20.-q}
%PACS, the Physics and Astronomy Classification Scheme.
%\keywords{Suggested keywords}%Use showkeys class option if keyword
                              %display desired
\maketitle
\section{Introduction}\label{sec:Intro}
Neutrinos have become a valuable alternative tool for probing gravity in various environments %\cite{DarkReview,MultiReview1} 
\cite{NeutrinoReview2022} thanks to their nonzero mass, their zero electric charge and their nonzero spin. Their nonzero mass makes them change their flavor as they propagate in space. First theoretically proposed in Refs.\,\cite{Pontecorvo1,Pontecorvo2}, and then theoretically predicted to exhibit a resonance inside matter in Refs.\,\cite{MSW1,MSW2}, these (experimentally well established  \cite{NeutrinoReview2022,SolarReview2022}) flavor oscillations of neutrinos are the favorite solution to the solar neutrino problem. On the other hand, neutrinos' zero electric charge and their weak interaction have prompted many authors to also investigate the effect of gravity on such oscillations, both within general relativity %\cite{Gasperini,Wudka,Ahluwalia1,Fornengo1,Bhattacharya,Cardall,Fornengo2,Remarks,InsideNO1,NeutrinoInCS,Lambiase1,HamiltonJacobi1,HamiltonJacobi2,InsideNO2,SuperNovaNO,Dvornikov,SpinNO,Geometric,Koutsoumbas,Blasone3,Capolupo,Xiu-Ju,Xin-Lian,Ren,JunPen,RenPan,Tao,Ren2,ChatelainVolpe,Petruzziello}, 
and within extended theories of gravity %\cite{Alexandre1,Alexandre2,Marletto,Torsion1,Torsion2,Chakraborty,Antonelli,Buoninfante,ConformalNO,ShortReview,Chakraborty1,Chakraborty2,Mandal,Arxiv2022}.
(see, \textit{e.g.}, the more recent Refs.\,\cite{NWPG,ConfNO,SNOMatter,Swami1,Swami2} and references therein.)

Being half-integer spin particles, however, neutrinos may exhibit yet another kind of oscillation beside the flavor oscillation. Indeed, like other spin particles, neutrinos possess a helicity and the latter may flip as they propagate in space. In fact, this possibility has been considered very early on \cite{Cisneros1970,Lim,Akhmedov} in an attempt to provide an alternative solution for the solar neutrino problem. For, as detectors respond only to left-handed neutrinos, any flip of neutrinos' helicity to right-handedness would indeed show up as a deficit in the expected number of detected neutrinos. This was the original motivation for considering spin oscillations since a nonzero magnetic moment of neutrinos could make the latter flip their helicity as they interact with a magnetic field along their path towards the detector. See the review \cite{Studenikin} and the compilation \cite{Book1994} and references therein for the early works, as well as Ref.\,\cite{Chukhnova2021} and references therein for more recent ones.

The possibility of spin flip caused by gravity has also been considered in the literature, first by studying arbitrary Dirac particles in Refs.\,\cite{Papini1991,Casini1994,Papini2004,Dolan2006}, and then by focusing on neutrinos in Refs.\,\cite{Dvornikov2006,Dvornikov2013,Alavi,Dvornikov2019,Dvornikov2020a,Dvornikov2020b,PRD2021}. In fact, it is well known that the general relativistic description of gravity entails that even in the absence of torque spinning bodies should exhibit a precession of their spin, caused either by the curvature of static spacetimes \cite{DeSitter} or by the frame dragging induced by a spinning gravitational source \cite{LensThirring}. Now, if this kind of spin precession also occurs for neutrinos a left-handed helicity neutrino could flip and become a right-handed one. This would then give rise to gravitationally induced spin oscillations of neutrinos analogous to the ones suggested in earlier works to occur due to magnetic fields. Furthermore, just as with the effect of gravity on neutrino flavor oscillations, the gravitationally induced spin oscillations of neutrinos could be studied within the framework of general relativity and within extended theories of gravity, including the much-studied screening models that offer an alternative explanation for dark energy. In this paper, we are interested in the latter category of gravitationally induced spin oscillations of neutrinos. 

The first motivation behind the present work is the fact that among the many interesting models proposed for the origin of dark energy \cite{DarkReview} are the so-called scalar-tensor theories of gravity and, more specifically, screening models in which a scalar field couples to matter via the spacetime metric. This special coupling to matter takes place in such a way that the unwanted ``fifth force'' interaction is screened (\textit{i.e.}, is rendered undetectable) at the Solar System level. Among the very well studied ones in this class of models are the chameleon model \cite{Chameleon,ChameleonIntroduction} and the symmetron model \cite{Symmetron}. The review \cite{ChameleonTests} offers a discussion of the various feasible experimental tests on such screening models.

The second motivation behind the present work is to consider the spin precession phenomenon within such screening models using a different and more general approach than the one adopted in Ref.\,\cite{PRD2021}. We first derive the general formula for the precession angular velocity as well as the probability for neutrinos helicity flip in arbitrary spacetimes and with an arbitrary screening scalar field. We then apply our results to the case of the spherically symmetric and static gravitational source described by the Schwarzschild metric. We consider the more realistic case of neutrinos deflected by a gravitational source along general paths, as opposed to the purely radial or circular paths. Our results show that, in contrast to what was reported in Ref.\,\cite{PRD2021}, the scalar field of chameleon-like and symmetron-like models has no effect on neutrinos' spin oscillations.

The remainder of this paper is divided into four sections as follows. In Sec.\,\ref{Sec:Precession}, after giving a brief introduction to the formalism used to describe spin dynamics in curved spacetimes, we derive the spin-precession angular velocity for neutrinos coupled to a general screening scalar field in an arbitrary spacetime. In Sec.\,\ref{Sec:Probability}, we use our results from Sec.\,\ref{Sec:Precession} to derive the probability formula for a neutrino to flip its helicity as it propagates in space. An application of the probability formula to chameleon-like and symmetron-like screening models will then be made in Sec.\,{\ref{Sec:ProbabilityApplications2} for the case of neutrinos deflected by a static spherically symmetric gravitational source. We conclude this paper with the brief section \ref{Sec:Conclusion} in which we summarise and discuss our main results.
%%%%%------%%%%%%%%
%%%%%------%%%%%%%%
%%%%%------%%%%%%%%
%%%%%------%%%%%%%%
%%%%%%%%%%%%%%%%%%%%%%%%%%%%%%%%%%%%%%%%%%%%%%%%%%%%%%%%%%
%%%%%%%%%%%%%%%%%%%%%%%%%%%%%%%%%%%%%%%%%%%%%%%%%%%%%%%%%
%%%%%%%%%%%%%%%%%%%%%%%%%%%%%%%%%%%%%%%%%%%%%%%%%%%%%%%%%

%%%%%%%%%%%%%%%%%%%%%%%%%%
%%%%%%%%%%%%%%%%%%%%%%%%%%
%%%%%%%%%%%%%%%%%%%%%%%%%%
%%%%%%%%%%%%%%%%%%%%%%%%%%
\section{Gravitationally and scalar-field induced spin precession}\label{Sec:Precession}
Thanks to the seminal work of Mathisson \cite{Mathisson} and the subsequent works of Papapetrou \cite{Papapetrou}
%Tulczyjew \cite{Tulczyjew}
and Dixon \cite{Dixon}, it became possible to describe the dynamics of any test-particle (including quantum particles) in curved spacetimes \cite{BakerReview} (see also Ref.\,\cite{Deriglazov} and references therein). It was shown in those works that spinning bodies in curved spacetime should exhibit not only non-geodesic motion, but also a spin precession. When keeping only the zeroth-order terms in spin on the right-hand side of the first set of Mathisson-Papapetrou-Dixon (MPD) equations, the deviation from geodesic motion of spinning bodies can safely be neglected. However, spin precession holds even if one restricts oneself to the zeroth-order terms in spin on the right-hand side of the second set of MPD equations. This fact allows one to seriously consider spin precession of quantum particles like neutrinos in curved spacetimes. In fact, while the extremely weak magnetic moment of neutrinos makes it hard to rely on spin precession caused by magnetic fields when dealing with the solar neutrino problem, the spacetime-induced spin precession occurs for any nonvanishing spin regardless of the value of the particle's magnetic moment. 

In this paper, we use (following Refs.\,\cite{Dvornikov2006,Dvornikov2013,Alavi,Dvornikov2020a,Dvornikov2020b,PRD2021}) the zeroth-order approximation of the second set of MPD equations to describe the dynamics of neutrinos' spin as they propagate in curved spacetime. In fact, although neutrinos in curved spacetime are governed by the curved-spacetime Dirac equation\footnote{We set throughout the paper $c=G=\hbar=1$, and we adopt the metric signature $(-,+,+,+)$.}, 
\begin{equation}
    \left(i\gamma^\mu\nabla_\mu-m\right)\psi=0,
\end{equation}
where $\gamma^\mu$ are the curved-spacetime gamma matrices and $\nabla_\mu$ is the spin covariant derivative, it was shown in Refs.\,\cite{Rudiger,Audretsch} that when using a Wentzel–Kramers–Brillouin (WKB) approximation to first order in $\hbar$, one does indeed recover the MPD equations from the Dirac equation. The MPD equations emerge from the Dirac equation by plugging into the latter the WKB ansatz, $\psi(x)=\exp\left[-\frac{i}{\hbar}S(x)\right]\sum_{n=0}^{\infty}\hbar^n\psi^{(n)}(x)$, and then equating to zero the coefficient of each power of $\hbar$. The result up to the power $\hbar^1$ is a Hamilton-Jacobi equation for the phase function $S(x)$ and a transport equation for the amplitude $\psi^{(0)}(x)$, from which emerge the MPD equations after using the Gordon decomposition of the Dirac probability current\footnote{Note that it is also possible to extract the MPD equations from the Dirac equation using an eikonal approximation \cite{CianfraniMontani1,CianfraniMontani2}. Yet, another way of describing spin dynamics of quantum spinning particles in curved spacetime is to use the curved-spacetime Dirac Hamiltonian in the Foldy-Wouthuysen representation \cite{2ndMethod}.}.

In Ref.\,\cite{Oancea}, such an approach was also applied to extract the MPD equations for charged particles in curved spacetime in the presence of a Maxwell field, as well as the MPD equations for massless particles. The latter case, on which we shall come back briefly below our result (\ref{SchwarSpinFlipProbaNumericalEvaluation}), is particularly interesting whenever one works within the approximation of massless neutrinos by neglecting their mass or by considering them within the ultra-relativistic regime.

The covariant description of spin (as required by a curved-spacetime representation of gravity) is achieved by conveniently introducing the spin four-vector $S^\mu$. In special relativity, the latter four-vector consists of the Lorentz boosted four-vector $(0,\textbf{S})$, where $\textbf{S}$ is the usual three-dimensional spin vector of the particle as measured by an observer at rest in the center-of-mass frame of the particle. In such a frame, the four-velocity $u^\mu={\rm d}x^\mu/{\rm d}\tau$ of the particle, where $\tau$ is an affine parameter usually chosen to be the proper time of the particle, is simply $(1,0,0,0)$. As such, the covariant identity $S^\mu u_\mu=0$ applies and it readily supplies us with the time-component $S^0$ of the spin vector $S^\mu$ in any other frame. However, the MPD equations are written in terms of the spin tensor $S^{\mu\nu}$ rather than the spin four-vector $S^\mu$. Indeed, at the zeroth order in spin on the right-hand side of the second MPD set of equations, the latter reduce to ${\rm D}S^{\mu\nu}/{\rm d}\tau=0$, where $\rm D$ is the covariant total derivative operator. Within the WKB approximation of the Dirac equation, the spin tensor emerges when one identifies $S^{\mu\nu}$ with $\frac{i\hbar}{4}(\bar{\psi}\psi)^{-1}\bar{\psi}[\gamma^\mu,\gamma^\nu]\psi$, where $\bar{\psi}$ is the Dirac adjoint of $\psi$ and $[\gamma^\mu,\gamma^\nu]$ is the commutator of the gamma matrices. 

Nevertheless, the spin vector $S^\mu$ can still be extracted from the spin tensor $S^{\mu\nu}$ in a similar manner as done in the flat-space special relativistic framework \cite{BMT}: $S^\mu=-\frac{1}{2m}\varepsilon^\mu_{\;\;\nu\rho\sigma}p^\nu S^{\rho\sigma}$ (also known as the Pauli–Lubanski pseudovector), where $\varepsilon^{\mu\nu\rho\sigma}$ is the totally antisymmetric Levi-Civita tensor and $p^\mu$ is the \textit{effective} four-momentum vector of the spinning particle. Although $p^\mu$ is not the usual four-momentum $mu^\mu$ of a particle of mass $m$ \cite{Mashhoon}, at the zeroth order in spin we do have the identity $p^\mu=mu^\mu$. On the other hand, since at the zeroth order the MPD equations for $p^\mu$ reduce to the geodesic equation ${\rm D}p^\mu/{\rm d}\tau=0$, we immediately see that at the zeroth order in spin we also have ${\rm D}S^\mu/{\rm d}\tau=0$. This is the equation that describes the dynamics of the spin four-vector as seen by an observer in the laboratory frame.

To properly describe spin precession, however, one needs to cast the equation ${\rm D}S^\mu/{\rm d}\tau=0$ back into the spinning particle's rest frame \cite{MTW}. This is done by using the projection $S^{\hat a}$ of the spin vector $S^\mu$ into the local tangent space of the moving particle thanks to the spacetime comoving vierbeins $e^{\hat a}_\mu$. These vierbeins are defined for any spacetime metric $g_{\mu\nu}$ by $\eta_{\hat{a}\hat{b}}e^{\hat a}_\mu e^{\hat b}_\nu=g_{\mu\nu}$, where $\eta_{\hat{a}\hat{b}}$ is the Minkowski metric\footnote{We use the first letters $(a,b,c)$ of the Latin alphabet to denote tangent-space indices while we reserve the Greek letters for curved-spacetime indices. The letters $(i,j,k)$ from the middle of the alphabet will be used to denote indices of the three-dimensional space}. The inverse comoving vierbeins $e^\mu_{\hat a}$ are defined by $e^{\hat a}_\mu e^\mu_{\hat b}=\delta^{\hat a}_{\hat b}$, where $\delta^{\hat a}_{\hat b}$ is the Kronecker delta symbol. Therefore, writing $S^{\hat a}=e^{\hat a}_\mu S^\mu$ and using ${\rm D}S^\mu/{\rm d}\tau=0$, we learn that the dynamics of $S^{\hat a}$ is described by the following equation,
\begin{equation}\label{S^a Equation}
\frac{{\rm d}S^{\hat a}}{{\rm d}\tau}=S^\mu\frac{{\rm D}e^{\hat a}_\mu}{{\rm d}\tau}.
\end{equation}

Next, recall that general spacetime vierbeins are related to the spin connection $\omega_\mu^{\,ab}$\footnote{This spin connection $\omega_\mu^{\,ab}$ is what is also denoted by $\gamma_{ba\mu}$ in the literature and called the Ricci rotation coefficients.} and the Christoffel symbols $\Gamma_{\mu\nu}^\lambda$ by $\omega_\mu^{\,\,ab}=-e^{\nu b}\partial_\mu e^a_\nu+e^{\nu b}\Gamma_{\mu\nu}^\lambda e^{a}_\lambda=-e^{\nu b}\nabla_\mu e^a_\nu$, where $\nabla_\mu$ denotes here the covariant derivative acting on tensors carrying curved-spacetime indices. Therefore, we have that $\nabla_\mu e^a_\nu=-e_{b\nu} \omega_\mu^{\,ab}$. Using this identity and the fact that $u^\mu={\rm d}x^\mu/{\rm d}\tau$ and $u^0={\rm d}t/{\rm d}\tau$, we extract from Eq.\,(\ref{S^a Equation}) the dynamics of the spin three-vector $\bf S$ (of components $S^{\hat i}$) in the rest frame of the particle in terms of the spin connection $\omega_\mu^{\,{\hat a}{\hat b}}$ as follows:
\begin{equation}\label{S^a Precession}
\frac{{\rm d}S^{\hat i}}{{\rm d}\tau}=-S_{\hat a}u^\mu\omega_\mu^{\,{\hat i}\hat{a}}.
\end{equation}
%Here, $u^{\hat0}=u^\mu e_\mu^{\hat 0}=1$ is the time-component of the four-velocity in the comoving frame. 
In Refs.\,\cite{Dvornikov2006,Dvornikov2013,Alavi,Dvornikov2020a,Dvornikov2020b,PRD2021}, the authors proceed from here by following Ref.\,\cite{Pomeranskii1997} (see also Ref.\,\cite{Pomeranskii2000}), where one relies on a similarity between Eq.\,(\ref{S^a Precession}) and the equation derived from the electrodynamics of relativistic particles \cite{QED}. However, to achieve our present aim, which is to investigate the effect on the spin precession of a particle coupled to a scalar field through the spacetime metric, it is very important to keep working solely with the spacetime tensors and not rely on analogies with flat-spacetime electrodynamics. 

Using the spin connection's antisymmetry, $\omega_\mu^{\,{\hat a}{\hat b}}=-\omega_\mu^{\,{\hat b}{\hat a}}$, and the fact that $S_{\hat 0}=0$, we extract from Eq.\,(\ref{S^a Precession}) the following dynamics for the three-dimensional spin vector $S^{\hat i}$:
\begin{equation}\label{S^iEquation}
    \frac{{\rm d}S^{\hat i}}{{\rm d}\tau}=-S_{\hat j}u^\mu\omega_\mu^{\,{\hat i}{\hat j}}\Leftrightarrow\frac{{\rm d}\textbf{S}}{{\rm d}\tau}={\bf\Omega}\times{\bf S}.
\end{equation}
In the second step we introduced the precession angular velocity vector $\bf\Omega$ which can be read off from the first equation to be \begin{equation}\label{OmegaVector}
    \Omega_{\hat i}=\tfrac{1}{2}\varepsilon_{{\hat i}{\hat j}{\hat k}}u^\mu\omega_{\mu}^{\,{\hat j}{\hat k}},
\end{equation}
where $\varepsilon_{{\hat i}{\hat j}{\hat k}}$ is the totally antisymmetric Levi-Civita symbol. The problem of finding the precession angular velocity reduces thus to finding the four velocity $u^\mu$ of the particle and the coefficients of the spin connection $\omega_{\mu}^{\;{\hat i}{\hat j}}$. The latter is, in turn, found as soon as one chooses the spacetime metric $g_{\mu\nu}$ from which one extracts the comoving vierbeins $e_\mu^{\hat a}$ corresponding to the four-velocity $u^\mu$.

Given that our problem is to deal with neutrinos coupled to the scalar field of the screening models, however, we cannot just use Eqs.\,(\ref{S^iEquation}) and (\ref{OmegaVector}) as they are. In fact, what is special about the screening models of interest to us here is that particles in those models are coupled to spacetime only thanks to a scalar field $\phi(x)$. As such, matter within those models behaves as if it evolves in a spacetime of metric $\tilde{g}_{\mu\nu}$ rather than in the original spacetime metric $g_{\mu\nu}$. Indeed, the metric $\tilde{g}_{\mu\nu}$ `seen' by matter in those models is the Weyl conformally rescaled metric $\tilde{g}_{\mu\nu}=A^2(\phi)g_{\mu\nu}$ \cite{Wald}, for some regular and nowhere vanishing functional $A(\phi)$ of the scalar field $\phi(x)$ of the models. 

This approach is motivated by the application of dilaton-field models to astrophysics and cosmology \cite{Damour1,Damour2}. The action of dilaton models is a scalar-tensor type action that can be chosen to have the form \cite{Damour1}:
\begin{equation}
    \mathcal{S}=\int\sqrt{-\tilde{g}}\,\left(\sigma\tilde{R}-\frac{\omega}{\sigma}\tilde{g}^{\mu\nu}\partial_\mu\sigma\partial_\nu\sigma\right)\,{\rm d}^4x+\int\mathcal{L}_m(\Psi_i,\tilde{g}_{\mu\nu})\,{\rm d}^4x,
\end{equation}
where $\sigma(x)$ is a scalar field, $\tilde{R}$ is the curvature scalar corresponding to the metric $\tilde{g}_{\mu\nu}$, $\omega$ is a Jordan-Fierz-Brans-Dicke parameter and $\Psi_i(x)$ are matter fields. The gravitational part of this action takes the more familiar Einstein-Hilbert form,
\begin{equation}
    \mathcal{S}=\frac{1}{16\pi}\int\sqrt{-g}\,\left(R-g^{\mu\nu}\partial_\mu\phi\partial_\nu\phi\right)\,{\rm d}^4x+\int\mathcal{L}_m(\Psi_i,\tilde{g}_{\mu\nu})\,{\rm d}^4x,
\end{equation}
after one performs the following metric and field  redefinitions: $\tilde{g}_{\mu\nu}=\exp[\phi/(\omega+\frac{3}{2})^{\frac{1}{2}}]\,g_{\mu\nu}$ and $\sigma=\frac{1}{16\pi}\exp[-\phi/(\omega+\frac{3}{2})^{\frac{1}{2}}]$, respectively. Matter fields $\Psi_i(x)$ in such models are thus described within any spacetime of metric $g_{\mu\nu}$ by replacing the matter Lagrangian $\mathcal{L}_m(\Psi_i,g_{\mu\nu})$ by the Lagrangian $\mathcal{L}_m(\Psi_i,\tilde{g}_{\mu\nu})$. For more generality, however, one allows for an arbitrary metric rescaling $\tilde{g}_{\mu\nu}=A^2(\phi)g_{\mu\nu}$. The functional $A(\phi)$ has the form $A(\phi)=\exp(\beta\phi)$ in the chameleon model \cite{Chameleon} and it has the form $A(\phi)=1+\beta\phi^2$ in the symmetron model \cite{Symmetron}, where $\beta$ is an arbitrary constant with the dimensions of an inverse mass. 

Therefore, to extract the modified equation describing spin precession within such screening models, one needs to figure out not only how is the spin connection altered, but also how does the spin vector $S^\mu$ get modified by the coupling with the scalar field and how is the four-velocity $u^\mu$ modified. Concerning the latter, it is straightforward to see that the effective four-velocity of the particle is $\tilde{u}^\mu={\rm d}x^\mu/{\rm d}\tilde{\tau}=A^{-1}(\phi)u^\mu$. For the spin connection, we also easily work out the new expression it takes under the metric rescaling by using its definition in terms of the covariant derivatives of the vierbeins (see Ref.\,\cite{SchroDi} for detailed steps of the calculation). We find,
\begin{equation}\label{RescaledConnection}
    \tilde{\omega}_\mu^{\,\,{\hat a}{\hat b}}=\omega_\mu^{\,\,{\hat a}{\hat b}}-\frac{A_{,\nu}}{A}\left(e^{\nu {\hat a}}e^{\hat b}_{\mu}-e^{\nu {\hat b}}e^{\hat a}_{\mu}\right).
\end{equation}
In deriving this expression, we used the fact that the spacetime vierbeins to which the particle couples are $\tilde{e}^{\hat a}_\mu=A(\phi)e^{\hat a}_\mu$. On the other hand, to find the effective spin vector $\tilde{S}^\mu$ of the particle, we need to recall the original definition of the spin tensor as being an integral associated to a pole-dipole test-particle approximation \cite{Dixon}. More explicitly, one first sets $\delta x^\mu=x^\mu-X^\mu$ for some arbitrary coordinates $X^\mu$ of a worldline chosen along the four-dimensional tube described by the motion of the particle. Then, one assumes that for all times the energy-momentum tensor $T^{\mu\nu}$ of the particle vanishes outside a sphere of radius $R\rightarrow0$ and centered at $X^i$ on some spacelike hypersurface $\Sigma$.  The integral giving the tensor $S^{\mu\nu}$ then reads
\begin{equation}\label{SpinTensor}
S^{\mu\nu}=\int\left(\delta x^\mu T^{\nu\rho}-\delta x^\nu T^{\mu\rho}\right){\rm d}\Sigma_\rho,
\end{equation}
where the integration is performed over all the spacelike hypersurface $\Sigma$. The effective spin tensor $\tilde{S}^{\mu\nu}$ is thus obtained by using the effective energy-momentum tensor $\tilde{T}^{\mu\nu}$ that does satisfy the conservation equation $\tilde{\nabla}_\nu \tilde{T}^{\mu\nu}=0$ \cite{ChameleonIntroduction}, where the covariant derivative $\tilde{\nabla}_\mu$ is associated to the rescaled metric $\tilde{g}_{\mu\nu}$. Under the rescaling of the spacetime metric, the usual definition of the energy-momentum tensor allows us to deduce that the energy-momentum tensor of a particle coupled to the scalar field that needs to be plugged into Eq.\,(\ref{SpinTensor}) is \begin{equation}\label{EMTransform}
 \tilde{T}^{\mu\nu}=\frac {2}{\sqrt{-\tilde{g}}}\frac{\delta \mathcal{L}_m(\Psi_i,\tilde{g}_{\mu\nu})}{\delta\tilde{g}_{\mu\nu}}=\frac {2}{\sqrt{-g}}\frac{\delta \mathcal{L}_m(\Psi_i,\tilde{g}_{\mu\nu})}{A^6(\phi)\,\delta g_{\mu\nu}}=\frac{T^{\mu\nu}}{A^6(\phi)}.   
\end{equation}
The energy-momentum tensor $\tilde{T}^{\mu\nu}$ is obtained by replacing inside the explicit expression of the familiar tensor $T^{\mu\nu}$ of a spinor field the metric $g_{\mu\nu}$ by its rescaled version $\tilde{g}_{\mu\nu}$. 
%Performing such a substitution inside the energy-momentum tensor of the fermion field $\psi$ \cite{QFT2}, we learn that
%\begin{equation}\label{DiracEM}
%    \tilde{T}^{\mu\nu}=\frac{i}{2}\bar{\psi}\tilde{\gamma}^{(\mu}\tilde{\nabla}^{\nu)}\psi-\frac{i}{2}[\tilde{\nabla}^{(\mu}\bar{\psi}]\tilde{\gamma}^{\nu)}\psi,
%\end{equation}
%where the round brackets stand for symmetrization in those indices, and $\tilde{\gamma}^\mu=A^{-1}(\phi)\gamma^\mu$ are the curved-spacetime gamma matrices associated to the metric $\tilde{g}_{\mu\nu}$. 

As the pole-dipole approximation on which integral (\ref{SpinTensor}) relies assumes that the dimensions of the spinning particle are very small compared with the characteristic length of the gravitational field under consideration \cite{Mashhoon}, the scalar field $\phi(x)$ can, to a very good approximation, be taken to be constant over the nonvanishing integration region in Eq.\,(\ref{SpinTensor}). Hence, we can move out of the integral all functional terms $A(\phi)$. This implies that we end up with $\tilde{S}^{\mu\nu}=A^{-2}(\phi)S^{\mu\nu}$. Note that this scaling of $S^{\mu\nu}$ is also what one arrives at using the spinor-based definition given above for $S^{\mu\nu}$ when recalling that $\gamma^\mu=e^\mu_{a}\gamma^{a}$, where  $\gamma^{a}$ are the constant gamma matrices, and that $[\tilde{\gamma}^\mu,\tilde{\gamma}^\nu]=\tilde{e}^\mu_{a}\tilde{e}^\nu_{b}[\gamma^{a},\gamma^{b}]=A^{-2}(\phi)[\gamma^\mu,\gamma^\nu]$. From this observation, we deduce that the effective spin vector $\tilde{S}^\mu$ of the coupled particle expressed fully in the metric $\tilde{g}_{\mu\nu}$ is related to the vector $S^\mu$ of the coupled particle expressed in the metric $g_{\mu\nu}$ by $\tilde{S}^\mu=A^{-1}(\phi)S^\mu$. Consequently, we conclude that we have $\tilde{S}^{\hat a}=\tilde{S}^\mu\tilde{e}_\mu^{\hat a}=S^{\hat a}$, which implies that the invariant squared spin magnitude $\tilde{S}^\mu\tilde{S}_\mu=s^2$ is, as it should, not altered by switching metrics either.

From these results, we finally deduce that the spin precession of the coupled particle is governed by Eqs.\,(\ref{S^iEquation}) and (\ref{OmegaVector}), where the proper time element ${\rm d}\tau$ should be replaced by ${\rm d}\tilde{\tau}$, the four-velocity $u^\mu$ should be replaced by $\tilde{u}^\mu$ and the spin connection should simply be replaced by the modified spin connection $\tilde{\omega}_\mu^{\,{\hat a}{\hat b}}$ as given by Eq.\,(\ref{RescaledConnection}). We thus conclude that the angular velocity vector of spin precession of neutrinos coupled to the scalar field reads
\begin{equation}\label{FinalAngularVelocity}
\tilde{\Omega}_{\hat i}=\varepsilon_{{\hat i}{\hat j}{\hat k}}\frac{u^\mu}{A}\left(\tfrac{1}{2}\,\omega_{\mu}^{\,\,{\hat j}{\hat k}}-\frac{A_{,\nu}}{A}e^{\nu{\hat j}}e_{\mu}^{\hat k}\right).
\end{equation}
Note that this final expression of the angular velocity is valid for any spacetime metric $g_{\mu\nu}$ and for any four-velocity $\tilde{u}^\mu=A^{-1}(\phi)u^\mu$ of the particle. The latter follows a spacetime geodesic given by $\tilde{D}\tilde{u}^\mu/{\rm d}\tilde{\tau}=0$, whereas a non-coupled particle follows a geodesic described by $D u^\mu/{\rm d}\tau=0$. In the next section, we shall drive a general formula for the spin-flip transition probability for any spin-precession angular velocity.

%%%%%%%%%%%%%%%%%%%%%%%%%%%%%%%%%
%%%%%%%%%%%%%%%%%%%%%%%%%%%%%%%%%
%%%%%%%%%%%%%%%%%%%%%%%%%%%%%%%%%
%%%%%%%%%%%%%%%%%%%%%%%%%%%%%%%%%
%%%%%%%%%%%%%%%%%%%%%%%%%%%%%%%%%
%%%%%%%%%%%%%%%%%%%%%%%%%%%%%%%%%
%%%%%%%%%%%%%%%%%%%%%%%%%%%%%%%%%
%%%%%%%%%%%%%%%%%%%%%%%%%%%%%%%%%
%%%%%%%%%%%%%%%%%%%%%%%%%%%%%%%%%
\section{Helicity flip probability}\label{Sec:Probability}
We can extract the spin-flip transition probability for any spin-precession angular velocity $\bf\Omega$ thanks to the effective Hamiltonian $H_{\rm eff}(\textbf{r})=\frac{1}{2}{\boldsymbol\sigma}.\tilde{\bf\Omega}$, where $\boldsymbol\sigma=(\sigma_1,\sigma_2,\sigma_3)$ are the three Pauli matrices. In the spherical coordinates $(r,\theta,\varphi)$ of interest to us here, we have $(\tilde{\Omega}_{1},\tilde{\Omega}_{2},\tilde{\Omega}_{3})=(\tilde{\Omega}_{\hat r},\tilde{\Omega}_{\hat\theta},\tilde{\Omega}_{\hat\varphi})$. When using the Pauli matrices in such coordinates \cite{Gravdi}, the effective time-independent Hamiltonian takes the following explicit form,
\begin{align}\label{SphericalHamiltonian}
H_{\rm eff}({\bf r})&=
\tfrac{1}{2}\tilde{\Omega}_1\begin{pmatrix}
\cos\theta & \sin\theta\, e^{-i\varphi}\\
\sin\theta\, e^{i\varphi} & -\cos\theta
\end{pmatrix}+
\tfrac{1}{2}\tilde{\Omega}_2\begin{pmatrix}
-\sin\theta & \cos\theta\, e^{-i\varphi}\\
\cos\theta\, e^{i\varphi} & \sin\theta
\end{pmatrix}\nonumber\\
&\quad+\tfrac{1}{2}\tilde{\Omega}_3\begin{pmatrix}
0 & -ie^{-i\varphi}\\
ie^{i\varphi} & 0
\end{pmatrix}.
\end{align}
Next, the general formula for the probability that a spin flips from an initial value $\ket{S_{\rm in}}$ at proper time $\tilde{\tau}_i$ to a final value $\ket{ S_{\rm fi}}$ at any other position at a later proper time $\tilde{\tau}_f$ is given by,
\begin{equation}\label{GeneralProbability}
\mathcal{P}(S_{\rm in}\rightarrow S_{\rm fi})=\left|\bra{S_{\rm fi}}{\bf T}\exp\left[-i\int_{\tilde{\tau}_i}^{\tilde{\tau}_f} H_{\rm eff}({\bf r})\,{\rm d}\tilde{\tau}\right]\ket{S_{\rm in}}\right|^2.
\end{equation}
In writing this expression, we introduced the usual time-ordering operator $\bf T$. Using that $H_{\rm eff}(\textbf{r})=\frac{1}{2}{\boldsymbol{\sigma}}.\tilde{{\boldsymbol{\Omega}}}(\textbf{r})$, as well as the well-known exponential identity $\exp(i\lambda {\bf n}.{\boldsymbol{\sigma}})=\cos\lambda+i{\bf n}.{\boldsymbol{\sigma}}\sin\lambda$, which is valid for any parameter $\lambda$ and for any unit vector $\bf n$, we can rewrite Eq.\,(\ref{GeneralProbability}) as follows 
\begin{align}\label{GeneralProbabilityExplicit}
 \mathcal{P}(S_{\rm in}\rightarrow S_{\rm fi})&=\left|\bra{S_{\rm fi}}\left[\cos\left(\int_{\tilde{\tau}_i}^{\tilde{\tau}_f}\frac{\tilde{\Omega}(\textbf{r})}{2}\,{\rm d}\tilde{\tau}\right)\right.\right.\nonumber\\
 &\quad\left.\left.-i\, \frac{\int_{\tilde{\tau}_i}^{\tilde{\tau}_f}\tilde{\boldsymbol{\Omega}}(\textbf{r}).\boldsymbol{\sigma}\,{\rm d}\tilde{\tau}}{\int_{\tilde{\tau}_i}^{\tilde{\tau}_f}\,\tilde{\Omega}(\textbf{r}){\rm d}\tilde{\tau}}\sin\left(\int_{\tilde{\tau}_i}^{\tilde{\tau}_f}\frac{\tilde{\Omega}(\textbf{r})}{2}{\rm d}\tilde{\tau}\right)\right]\ket{S_{\rm in}}\right|^2, 
\end{align}
where $\tilde{\Omega}(\textbf{r})$ stands for the magnitude of the angular velocity vector at the position $\textbf{r}$ of the particle.
%%%%%%%%%%%%%%%%%%%%%%%%%%%%%%%%%%%
%%%%%%%%%%%%%%%%%%%%%%%%%%%%%%%%%%%%
%%%%%%%%%%%%%%%%%%%%%%%%%%%%%%%%%%%%
%%%%%%%%%%%%%%%%%%%%%%%%%%%%%%%%%%%
%%%%%%%%%%%%%%%%%%%%%%%%%%%%%%%%%%%%
%%%%%%%%%%%%%%%%%%%%%%%%%%%%%%%%%%%%

What we are interested in here, is the case of left-handed neutrinos deflected by a spherical massive gravitational source. In what follows we assume, for definiteness and simplicity, that the motion of the particle is along the equatorial plane $\theta=\frac{\pi}{2}$ of the spherical coordinates. Suppose the particle came from infinity and got deflected counterclockwise by the gravitational source. We therefore take the initial spin of the particle to be anti-parallel to the direction of its initial motion, such that the initial spin state satisfies ${\bf n}.{\boldsymbol{\sigma}}\ket{S_{\rm in}}=-\ket{S_{\rm in}}$, where $\bf n$ is the unit vector along the direction of motion of the particle. The initial orthonormalized spin state of the neutrino then reads 
\begin{equation}\label{NegativeHelicityAzimuthalSpin}
\ket{S_{\rm in}}=
     \frac{1}{\sqrt{2}}\begin{pmatrix}
     -1 \\
    \eta e^{i\varphi}
\end{pmatrix},    
\end{equation}
where $\eta=(u^1+iu^3)/\sqrt{(u^1)^2+(u^3)^2}$. We are looking for the probability that the helicity becomes positive, \textit{i.e.}, for the spin to be found aligned along the positive direction of motion at a detection point very far from the gravitational source. Such a final spin state has the form,
\begin{equation}\label{PositiveHelicityAzimuthalSpin}
\ket{S_{\rm fi}}=\frac{1}{\sqrt{2}}\begin{pmatrix}
     1\\
    \eta e^{i\phi}
\end{pmatrix}.
\end{equation}
Plugging these two spin states into the general formula (\ref{GeneralProbabilityExplicit}), we find, after taking account of the Hamiltonian (\ref{SphericalHamiltonian}), the following probability for a left-handed neutrino $\ket{\nu_L}$ to become a right-handed neutrino $\ket{\nu_R}$:
\begin{align}\label{OrbitingSpinProbability}
&\mathcal{P}(\ket{\nu_L}\rightarrow \ket{\nu_R})=\left[\int_{\tilde{\tau}_i}^{\tilde{\tau}_f}\,\tilde{\Omega}(\textbf{r})\,{\rm d}\tilde{\tau}\right]^{-2}\sin^2\left[\tfrac{1}{2}\int_{\tilde{\tau}_i}^{\tilde{\tau}_f}\tilde{\Omega}(\textbf{r})\,{\rm d}\tilde{\tau}\right]\nonumber\\
&\quad\times\left(\left[\int_{\tilde{\tau}_i}^{\tilde{\tau}_f}\,\mathfrak{Im}\!\left(\eta\left[\tilde{\Omega}_{\hat 1}(\textbf{r})-i\tilde{\Omega}_{\hat 3}(\textbf{r})\right]\right){\rm d}\tilde{\tau}\right]^2+\left[\int_{\tilde{\tau}_i}^{\tilde{\tau}_f}\,\tilde{\Omega}_{\hat 2}(\textbf{r})\,{\rm d}\tilde{\tau}\right]^2\right).
\end{align}
Here, $\mathfrak{Im}(\eta[\tilde{\Omega}_{\hat 1}(\textbf{r})-i\tilde{\Omega}_{\hat 3}(\textbf{r})])$ stands for the imaginary part of the complex expression $\eta[\tilde{\Omega}_{\hat 1}(\textbf{r})-i\tilde{\Omega}_{\hat 3}(\textbf{r})]$. In the next section we apply this general formula to neutrinos experiencing a spin precession of angular velocity (\ref{FinalAngularVelocity}) as they get deflected by a static and spherical gravitational source. 

Note that, unlike what one does when studying neutrino flavor oscillations under the effect of gravity where one works in the mass eigenstates basis $(\ket{\nu_1},\ket{\nu_2},\ket{\nu_3})$, we work here in the flavor basis where $\ket{\nu_L}$ and $\ket{\nu_R}$ simply denote the helicity of the specific flavor we are interested in; either an electron-neutrino $\ket{\nu_e}$, or a muon-neutrino $\ket{\nu_\mu}$, or a tau-neutrino $\ket{\nu_\tau}$.

%%%%%------%%%%%%%%
%%%%%------%%%%%%%%
%%%%%------%%%%%%%%
%%%%%------%%%%%%%%
%%%%%%%%%%%%%%%%%%%%%%%%%%%%
%%%%%%%%%%%%%%%%%%%%%%%%%%%%
%%%%%%%%%%%%%%%%%%%%%%%%%%%%
%%%%%%%%%%%%%%%%%%%%%%%%%%%%
%%%%%%%%%%%%%%%%%%%%%%%%%%%%%%%%
%%%%%%%%%%%%%%%%%%%%%%%%%%%%%%%%
%%%%%%%%%%%%%%%%%%%%%%%%%%%%%%%%
%%%%%%%%%%%%%%%%%%%%%%%%%%%%%%%%
%%%%%------%%%%%%%%
%%%%%------%%%%%%%%
%%%%%------%%%%%%%%
%%%%%------%%%%%%%%
%%%%%%%%%%%%%%%%%%%%%%%%%%%%
%%%%%%%%%%%%%%%%%%%%%%%%%%%%
%%%%%%%%%%%%%%%%%%%%%%%%%%%%
%%%%%%%%%%%%%%%%%%%%%%%%%%%%
\section{Neutrinos deflected by a static spherical gravitational source}\label{Sec:ProbabilityApplications2}

It is easy to see from formula (\ref{FinalAngularVelocity}) that for a static spherical gravitational source, only orbiting neutrinos would display a nonzero spin-precession angular velocity. The reason is that a spherical gravitational source would also give rise only to a radially-varying scalar field $\phi$, and hence only the component $A_{,r}$ of the gradient of $A(\phi)$ is nonvanishing in Eq.\,(\ref{FinalAngularVelocity}). Therefore, we assume here \textit{deflected} neutrinos (as opposed to radially-propagating neutrinos) around a static spherical mass $M$, using the Schwarzschild metric given by
\begin{equation}\label{Schwarzschild}
    {\rm d}s^2=-\left(1-\frac{2M}{r}\right){\rm d}t^2+\left(1-\frac{2M}{r}\right)^{-1}{\rm d}r^2+r^2\left({\rm d}\theta^2+\sin^2\theta\,{\rm d}\varphi^2\right).
\end{equation}
Next, we assume the particle starts moving from infinity with an energy $\tilde{\mathcal{E}}=-\tilde{g}_{00}\tilde{p}^0=-\tilde{g}_{00}m\tilde{u}^0$ before getting closer to the static mass and getting deflected around the latter in the equatorial plane. Note that, as opposed to what one does when studying flavor oscillations, the mass $m$ used here is that of a flavor state rather than one of the three different masses associated to the mass eigenstates. This energy of the neutrino is again related at each position $\bf r$ to the conserved energy $\mathcal{E}$ at infinity of a neutrino not coupled to the scalar field by $\tilde{\mathcal{E}}=A(\phi)\mathcal{E}$. Also, we denote the conserved orbital angular momentum by $L$. The particle follows then a constant-$L$ and constant-$\mathcal{E}$ (rather than constant-$\tilde{L}$ and constant-$\tilde{\mathcal{E}}$) path with four-velocity $\tilde{u}^\mu=A^{-1}(\phi)u^\mu$, where $u^\mu$ is given by \cite{Hartle}:
\begin{equation}\label{4VeloOrbitingSchwar}
    u^\mu=\left(\frac{e}{\Delta}, \pm\sqrt{e^2-\Sigma\Delta}, 0, \frac{\ell}{r^2}\right).
\end{equation}
For convenience, we set here $\Delta=1-2M/r$, $\Sigma=1+\ell^2/r^2$, $e=\mathcal{E}/m$ and $\ell=L/m$. The $(+)/(-)$ signs stand, respectively, for neutrinos moving outward/inward relative to the central mass. On the other hand, with the help of this four-velocity we easily compute the corresponding vierbeins and the corresponding spin connection coefficients displayed in Appendix \ref{Sec:AppB}. Plugging the vierbeins (\ref{AppOrbitSchwarVierbeins}) and (\ref{AppInverseOrbitSchwarVierbeins}) and the nonzero spin connection coefficients (\ref{AppOrbitSchwarSpinConnect}) into formula (\ref{FinalAngularVelocity}), we find the following expression for the single nonvanishing component of the spin-precession angular velocity:
\begin{equation}\label{SchwarOrbitAngularVelo}
\tilde{\Omega}_{\hat 2}=\pm\frac{e\ell}{r^2A\Sigma}.
\end{equation}
Again, the ($+$)/($-$) signs stand for particles moving outward/inward relative to the central mass. We notice that no dependence on the radial gradient of the scalar field $\phi$ appears in this formula. 

Next, we also need to convert the ${\rm d}\tilde{\tau}$ in the integrals in Eq.\,(\ref{OrbitingSpinProbability}) into a ${\rm d}r$. Referring to the four-velocity (\ref{4VeloOrbitingSchwar}), we deduce that the proper time of the neutrinos is governed by the following differential equation relating the proper time element ${\rm d}\tilde{\tau}$ to the coordinate radius element ${\rm d}r$:
\begin{equation}\label{Schwardtdr}
{\rm d}\tilde{\tau}=\pm\frac{A\,{\rm d}r}{\sqrt{e^2-\Sigma\Delta}}.   
\end{equation}
Here, we used the ($\pm$) sign to guarantee a positive proper time, for ${\rm d}r$ is positive for outward moving particles and negative for inward moving ones. The coordinate radius $r$ decreases along the trajectory of the neutrino until it reaches its minimum value $r_0$ at the turning point where ${\rm d}r/{\rm d}\tau=0$, and starts increasing as the particle continues its journey away from the gravitational source. The coordinate radius $r=r_0$ of the closest approach of the particle to the gravitational source is thus found by solving the equation $\Sigma\Delta=e^2$. To evaluate the probability integrals over $r$ in Eq.\,(\ref{OrbitingSpinProbability}), we need to pick up the same relative signs from expressions (\ref{SchwarOrbitAngularVelo}) and (\ref{Schwardtdr}). This amounts then to integrating over $r$ from $r=r_0$ to $r=\infty$ using an overall ($+$) sign, and multiplying by a factor of $2$ the resulting integral.

Plugging Eq.\,(\ref{SchwarOrbitAngularVelo}) into Eq.\,(\ref{OrbitingSpinProbability}), after setting $\tilde{\Omega}_{\hat1}=\tilde{\Omega}_{\hat3}=0$ and making use of identity (\ref{Schwardtdr}), we find
\begin{align}\label{SchwarSpinFlipProbaIntegral}
\mathcal{P}(\ket{\nu_L}\rightarrow \ket{\nu_R})&=\sin^2\left(\int_{r_0}^{\infty}\frac{2e\ell}{r^2\Sigma\sqrt{e^2-\Sigma\Delta}}{\rm d}r\right).
\end{align}
This result shows no trace of the scalar field $\phi$. The latter disappeared from the integral after the factor $A(\phi)$ in the denominator of the angular frequency $\tilde{\Omega}_{\hat 2}$ got canceled by the same factor in the numerator of expression (\ref{Schwardtdr}) of the proper time. All one extracts then from formula (\ref{SchwarSpinFlipProbaIntegral}) is the spin-flip probability of neutrinos moving within the purely gravitational field of the central mass, as if the particles were completely decoupled from the scalar field. Thus, the spin-flip probability is not affected by the scalar field of the screening models considered here. Nevertheless, the result (\ref{SchwarSpinFlipProbaIntegral}) is general enough and is thus interesting for its own sake.

In fact, formula (\ref{SchwarSpinFlipProbaIntegral}) is valid for any energy $e$ per unit mass of the neutrinos, for any angular momentum $\ell$ per unit mass of the latter, for any possible closest approach $r_0$ of the particles, and for any mass $M$ of the gravitational source. Unfortunately, no analytic expression of the integral in Eq.\,(\ref{SchwarSpinFlipProbaIntegral}) could be found for such a general set of parameters. Only a numerical integration can be performed once one chooses specific values for each of those four parameters. 

For that purpose, we shall assume $e^2-1\approx e^2$, which is a good approximation for the neutrinos of interest to us here as the latter are necessarily coming from interstellar high-energy sources before they get deflected by a gravitational source located on their path. Further, we set $x=\ell/r$ and make use of the usual definition of the impact parameter $b=\ell/\sqrt{e^2-1}\approx\ell/e$ that describes the trajectory of deflected particles by the gravitational field of a central mass. Therefore, we have $x\approx be/r$. For definiteness,  we choose $r_0=b$, so that $x_0=be/r_0\approx e$, and we choose $e=10^7$ for a typical parameter of supernovae neutrinos (see, {\it{e.g.}}, Ref.\,\cite{NeutrinoBook2} and the references therein). Allowing for large-curvature scenarios, we choose the impact parameter $b$ to be only twice the Schwarzschild radius of the gravitational source, so that by introducing the dimensionless parameter $\mu=M/\ell$, we have $\mu\approx M/be=5\times10^{-8}$. The integral in Eq.\,(\ref{SchwarSpinFlipProbaIntegral}) can then be evaluated numerically, yielding  
\begin{align}\label{SchwarSpinFlipProbaNumericalEvaluation}
\mathcal{P}(\ket{\nu_L}\!\rightarrow\! \ket{\nu_R})&=\sin^2\left(\int_0^{x_0}\!\!\!\!\frac{2e}{\left(1+x^2\right)\!\!\sqrt{e^2-\left(1+x^2\right)\left(1-2\mu x\right)}}{\rm d}x\right)\nonumber\\
&\approx\sin^2\pi.
\end{align}
This result shows that the spin-flip probability vanishes for such high-energy, {\it{i.e.}}, ultra-relativistic neutrinos. This conclusion is in agreement with what has been found in Ref.\,\cite{Dolan2006}. This conclusion is also in agreement with what the MPD equations extracted from the Dirac equation imply for massless particles \cite{Oancea}: static curved spacetime does not induce spin precession on massless particles. 

Since our main goal here is rather to study the dependence of the spin-flip probability on the variation of those four parameters listed above, we shall focus now on the more analytically tractable small-curvature case for which the approximation $M/r\ll1$ applies.
%%%%%%%%%%%%%%%%%%%%%%
%%%%%%%%%%%%%%%%%%%%%%
%%%%%%%%%%%%%%%%%%%%%%
%%%%%%%%%%%%%%%%%%%%%%
%%%%%%%%%%%%%%%%%%%%%%
%%%%%%%%%%%%%%%%%%%%%%
%%%%%%%%%%%%%%%%%%%%%%
%%%%%%%%%%%%%%%%%%%%%%
%%%%%%%%%%%%%%%%%%%%%%
For $M/r\ll1$, we may expand the integrand in Eq.\,(\ref{SchwarSpinFlipProbaIntegral}) up to the leading power in $M/r$. Performing again a change of variables by setting $x=\ell/r$, and introducing the dimensionless parameter $\mu=M/\ell$, formula (\ref{SchwarSpinFlipProbaIntegral}) takes the following form,
\begin{align}\label{SchwarSpinFlipProbaApprox}
\mathcal{P}(\ket{\nu_L}\rightarrow \ket{\nu_R})&\approx\sin^2\left(\int_{0}^{x_0}\frac{2e(1+\mu x)}{(1+x^2)\sqrt{e^2+2\mu e^2x-x^2}}{\rm d}x\right),
\end{align}
where we set $x_0=\ell/r_0$. The integral in this equation can be analytically evaluated, yielding
\begin{widetext}
\begin{multline}\label{EvaluatedSchwarSpinFlipProbaApprox}
\mathcal{P}(\ket{\nu_L}\rightarrow \ket{\nu_R})\approx\sin^2\left\{-\frac{\pi \eta}{2} \left(\frac{\sqrt{1+\chi^2}+1}{\sqrt{1+\chi^2}}\right)^{\!\frac{1}{2}}\left(2\sqrt{2}-\tan^{-1}\chi\right)\right.\\
\left.+\frac{\eta}{2\sqrt{2}}\left(\frac{\sqrt{1+\chi^2}-1}{\sqrt{1+\chi^2}}\right)^{\frac{1}{2}}\ln\left[\frac{(1+\eta^2)\sqrt{1+\chi^2+\sqrt{1+\chi^2}}+\sqrt{2}\eta\left(1+\sqrt{1+\chi^2}\right)}{(1+\eta^2)\sqrt{1+\chi^2+\sqrt{1+\chi^2}}-\sqrt{2}\eta\left(1+\sqrt{1+\chi^2}\right)}\right]\right\},
\end{multline}
with
\begin{equation}\label{EtaChi}
    \eta=\left(\frac{1+\mu^2}{\sqrt{1+4\mu^2}}\right)^{\frac{1}{2}},\qquad \chi=\frac{2\mu^3}{1+3\mu^2}. 
\end{equation}
\end{widetext}
The spin-flip probability (\ref{EvaluatedSchwarSpinFlipProbaApprox}) is thus expressed solely in terms of the parameter $\mu=M/\ell$. It has no dependence on the radial coordinate $r_0$ of the neutrinos' closest approach to the central mass. The disappearance of $r_0$ from Eq.\,(\ref{EvaluatedSchwarSpinFlipProbaApprox}) came about due to the fact that at this order of the expansion in $M/r$, the denominator in Eq.\,(\ref{SchwarSpinFlipProbaApprox}) vanishes at $x=x_0$.

In order to examine the dependence of the probability (\ref{EvaluatedSchwarSpinFlipProbaApprox}) on the variation of the energy of the neutrinos and on the variation of the shape of their deflected paths, we replace $\mu$ by $M/be$ in Eq.\,(\ref{EtaChi}). The latter yields then, 
\begin{equation}\label{EtaChiWithb}
    \eta=\left(\frac{b^2e^2+M^2}{be\sqrt{b^2e^2+4M^2}}\right)^{\!\frac{1}{2}},\qquad \chi=\frac{2M^3}{be(b^2e^2+3M^2)}. 
\end{equation}
Using these expressions of $\eta$ and $\chi$, we can not only track down the dependence of the probability (\ref{EvaluatedSchwarSpinFlipProbaApprox}) on the variation of the parameters $e$ and $b$ of the neutrinos, but we can also find out about the dependence of the probability on the variation of the mass $M$ of the gravitational source. In fact, let us consider typical values of the energy $e$ per unit mass of the neutrinos by allowing $e$ to be as low as $e\sim10^5$ for solar neutrinos and as high as $e\sim10^7$ for supernovae neutrinos. We also consider an impact parameter $b$ as small as a few dozens of $M_\odot$, that might occur for deflections around compact neutron stars. Indeed, for such impact parameters around such compact stars we may keep the second order in $M/be$ in the parameter $\eta$, discarding only the parameter $\chi$ for being third-order in $M/be$. From Eq.\,(\ref{EtaChiWithb}), we then have
\begin{equation}\label{EtaChiFirstOrder}    
\eta\approx1-\frac{M^2}{2b^2e^2},\qquad \chi\approx0. 
\end{equation}
Note that at the first order in $M/be$, we have $\eta=1$ and $\chi=0$. Plugging these two values into Eq.\,(\ref{EvaluatedSchwarSpinFlipProbaApprox}) we recover the vanishing spin-flip probability. On the other hand, plugging the approximations (\ref{EtaChiFirstOrder}) into Eq.\,(\ref{EvaluatedSchwarSpinFlipProbaApprox}), the latter takes the form,
\begin{equation}\label{SchwarSpinFlipProbaApproximated}
\mathcal{P}(\ket{\nu_L}\rightarrow \ket{\nu_R})\approx\frac{\pi^2 M^4}{b^4e^4}.
\end{equation}
Since the impact parameter cannot be smaller than the gravitational radius of the central mass, we clearly see from Eq.\,(\ref{SchwarSpinFlipProbaApproximated}) that although the result increases like the fourth power of the central mass, the spin-flip probability is suppressed as the inverse fourth power of the energy per unit mass of the neutrinos for typical neutrino energies and for any realistic impact parameter $b$. 
%%%%%%%%%%%%%%%%%%%%%%%%%%%%%%%%%%%%%
%%%%%%%%%%%%%%%%%%%%%%%%%%%%%%%%%%%%%
%%%%%%%%%%%%%%%%%%%%%%%%%%%%%%%%%%%%%
%%%%%%%%%%%%%%%%%%%%%%%%%%%%%%%%%%%%%
%%%%%%%%%%%%%%%%%%%%%%%%%%%%%%%%%%%%%
%%%%%%%%%%%%%%%%%%%%%%%%%%%%%%%%%%%%%
%%%%%------%%%%%%%%
%%%%%------%%%%%%%%
%%%%%------%%%%%%%%
%%%%%------%%%%%%%%
%%%%%%%%%%%%%%%%%%%%%%%%%%
%%%%%%%%%%%%%%%%%%%%%%%%%%
%%%%%%%%%%%%%%%%%%%%%%%%%%
%%%%%%%%%%%%%%%%%%%%%%%%%%
%%%%%%%%%%%%%%%%%%%%%%%%%%
%%%%%%%%%%%%%%%%%%%%%%%%%%
%%%%%%%%%%%%%%%%%%%%%%%%%%
%%%%%%%%%%%%%%%%%%%%%%%%%%
%%%%%%%%%%%%%%%%%%%%%%%%%%
%%%%%%%%%%%%%%%%%%%%%%%%%%
%%%%%%%%%%%%%%%%%%%%%%%%%%
%%%%%%%%%%%%%%%%%%%%%%%%%%
%%%%%%%%%%%%%%%%%%%%%%%%%%
%%%%%%%%%%%%%%%%%%%%%%%%%%
%%%%%------%%%%%%%%
%%%%%------%%%%%%%%
%%%%%------%%%%%%%%
%%%%%------%%%%%%%%
%%%%%%%%%%%%%%%%%%%%%%%%%%%%
%%%%%%%%%%%%%%%%%%%%%%%%%%%%
%%%%%%%%%%%%%%%%%%%%%%%%%%%%
%%%%%%%%%%%%%%%%%%%%%%%%%%%%
\section{Summary and conclusion}\label{Sec:Conclusion}
We considered neutrinos spin oscillations within curved spacetime when the neutrinos are coupled to the scalar field of chameleon-like and symmetron-like screening models. We first derived the general expression of the spin-flip probability in the flavor basis by relying on the MPD equations that describe classical spin precession under the effect of gravity. We then applied the formula to extract the corresponding helicity-flip probability for the case of neutrinos deflected in the equatorial plane of a static and spherically symmetric gravitational source. Our result is valid for any shape of the deflected neutrinos paths and it is given in terms of the impact parameter of the particles, their energy per unit mass and the coordinate radius of their closest approach.

What is remarkable about our result is not only the absence of any dependence of the spin-flip probability on the radial variation of the scalar field, but also the absence of the scalar field altogether. This means that, although the scalar field does have an effect on the neutrinos' spin precession in the comoving frame, as shown by the spin-precession angular velocity (\ref{SchwarOrbitAngularVelo}), the spin-flip probability is not affected by the coupling of the neutrinos to the scalar field.

The reason why the spin-precession angular velocity is affected but the spin-flip probability is not is due to the peculiar coupling we considered here of the screening models with matter. In fact, the scalar field $\phi$ of such models couples to neutrinos only through the spacetime metric by Weyl-rescaling the latter. As such, both the proper time of the particle and its spin-precession angular velocity in the comoving frame are oppositely affected. The proper time is affected by being multiplied by the functional $A(\phi)$ of the scalar field, but the angular velocity gets divided by the functional $A(\phi)$. The functional $A(\phi)$ thus cancels out of the probability integral. Another way of understanding this result, is to recall that the conformal rescaling of the metric caused by the scalar field preserves angles, but rescales distances and time. It is therefore no wonder that the total precession angle and the corresponding probability for its occurrence both remain unaffected.  

Even though our result showed no effect of the scalar field on the helicity-flip probability, the generality of the latter's final expression allowed us to examine closer the effect of pure gravity in the absence of any coupling. In the process, we recovered the vanishing of the probability for high-energy neutrinos with any realistic impact parameter. We also extracted a closed form showing the dependence of the probability on the energy per unit mass of the neutrinos, on the impact parameter of the latter, and on the mass of the gravitational source.     

It is worth emphasising here that our present study focused solely on neutrinos deflected by a static and spherically symmetric gravitational source. Nevertheless, the physical interpretation we provided for our present result is general enough to bring into light the main insights and to encompass arbitrary physical situations of the interaction of the spin with the gravitational field of the source. Still, an extension of this work to include general motions of neutrinos and/or rotating gravitational sources will be provided elsewhere. Furthermore, to examine any eventual fundamental difference between the use of the mass eigenstates basis and the use of the flavor basis another study based on the Foldy-Wouthuysen representation as opposed to the MPD equations, as well as a study based purely on the WKB approximation for deriving the MPD equations from the Dirac equation, are both necessary and will also be carried out elsewhere.

%%%%%%%%%%%%%%%%%%%%%%%%%%%%%%%%%%
%%%%%%%%%%%%%%%%%%%%%%%%%%%%%%%%%%
%%%%%%%%%%%%%%%%%%%%%%%%%%%%%%%%%%
\section*{Acknowledgments}
The authors are grateful to the anonymous referees for their constructive comments that helped improve our presentation. This work was supported by the Natural Sciences and Engineering Research Council of Canada (NSERC) Discovery Grant No. RGPIN-2017-05388; and by the Fonds de Recherche du Québec - Nature et Technologies (FRQNT). PS acknowledges support from Bishop's University Research Assistantship award.

%%%%%%%%%%%%%%%%%%%%%%%%%%%%%%%%
%%%%%%%%%%%%%%%%%%%%%%%%%%%%%%%%
%%%%%%%%%%%%%%%%%%%%%%%%%%%%%%%%
\appendix
%%%%%------%%%%%%%%
%%%%%------%%%%%%%%
%%%%%------%%%%%%%%
%%%%%------%%%%%%%%
%%%%%------%%%%%%%%
%%%%%------%%%%%%%%
\section{The comoving vierbeins and the spin connection}\label{Sec:AppB}
We display here the comoving vierbeins one builds from the Schwarzschild metric (\ref{Schwarzschild}) and the four-velocity (\ref{4VeloOrbitingSchwar}) of a deflected neutrino around a static and spherically symmetric gravitational source. We set $\mathcal{E}/m=e$ and $L/m=\ell$, where $\mathcal{E}$ is the conserved energy, $m$ is the mass of the neutrino and $L$ is the conserved angular momentum of the latter. We also set $\Delta=1-2M/r$ and $\Sigma=1+\ell^2/r^2$. Then, the resulting vierbeins read,
\begin{align}\label{AppOrbitSchwarVierbeins}
    e_{\hat 0}^\mu&=\left(\frac{e}{\Delta}, \pm\sqrt{e^2-\Sigma\Delta}, 0, \frac{\ell}{r^2}\right),\nonumber\\
e_{\hat1}^\mu&=\left(\frac{\sqrt{e^2-\Sigma\Delta}}{\sqrt{\Sigma}\Delta},\pm\frac{e}{\sqrt{\Sigma}},0,0\right),\nonumber\\
e_{\hat 2}^\mu&=\left(0,0,\pm\frac{1}{r},0\right),\nonumber\\
e_{\hat 3}^\mu&=\left(\frac{e\ell}{r\sqrt{\Sigma}\Delta},\pm\frac{\ell\sqrt{e^2-\Sigma\Delta}}{r\sqrt{\Sigma}},0,\frac{\sqrt{\Sigma}}{r}\right);
\end{align}
and their inverses $e^{\hat a}_\mu$ are
\begin{align}\label{AppInverseOrbitSchwarVierbeins}
    e^{\hat 0}_\mu&=\left(e,\mp\frac{\sqrt{e^2-\Sigma\Delta}}{\Delta}, 0, -\ell\right),\nonumber\\ 
    e^{\hat 1}_\mu&=\left(-\frac{\sqrt{e^2-\Sigma\Delta}}{\sqrt{\Sigma}},\pm\frac{e}{\sqrt{\Sigma}\Delta},0,0\right),\nonumber\\ e^{\hat 2}_\mu&=\left(0,0,\pm r,0\right),\nonumber\\
    e^{\hat 3}_\mu&=\left(-\frac{e\ell}{r\sqrt{\Sigma}},\pm\frac{\ell\sqrt{e^2-\Sigma\Delta}}{r\sqrt{\Sigma}\Delta},0,r\sqrt{\Sigma}\right).
\end{align}
On the other hand, the nonzero Christoffel symbols one extracts from the Schwarzschild metric are:
\begin{align}
\Gamma_{00}^1&=\frac{M\Delta}{r^2},\;\;\,\qquad\quad \Gamma_{01}^0=\frac{M}{r^2\Delta},\;\qquad \Gamma_{11}^1=-\frac{M}{r^2\Delta},\nonumber\\
\Gamma_{12}^2&=\Gamma_{13}^3=\frac{1}{r},\;\quad\quad
\Gamma_{22}^1=-r\Delta,\;\qquad \Gamma_{23}^3=\cot\theta,\nonumber\\
\Gamma_{33}^1&=-r\sin^2\theta\,\Delta,\;\quad \Gamma_{33}^2=-\sin\theta\cos\theta.
\end{align}
From these expressions, we compute the following relevant nonzero coefficients of the spin connection to be
\begin{align}\label{AppOrbitSchwarSpinConnect}
&\omega_{0}^{\,{\hat1}{\hat3}}=\pm\frac{M\ell}{r^3},\nonumber\\
&\omega_{1}^{\,{\hat1}{\hat3}}=\frac{e\ell}{\Sigma\Delta\sqrt{e^2-\Sigma\Delta}}\left(\frac{\ell^2}{r^4}-\frac{M}{r^3}-\frac{3M\ell^2}{r^5}\right),\nonumber\\
&\omega_{3}^{\,{\hat1}{\hat3}}=\mp e.
\end{align}
%%%%%------%%%%%%%%
%%%%%------%%%%%%%%
%%%%%------%%%%%%%%
%%%%%------%%%%%%%%
%%%%%------%%%%%%%%
%%%%%------%%%%%%%%

%%%%%------%%%%%%%%
%%%%%------%%%%%%%%
%%%%%------%%%%%%%%
%%%%%------%%%%%%%%
%%%%%------%%%%%%%%
%%%%%------%%%%%%%%
%%%%%------%%%%%%%%
%%%%%------%%%%%%%%
%%%%%------%%%%%%%%
%%%%%------%%%%%%%%


\begin{thebibliography}{}

\bibitem{NeutrinoReview2022} M.\,S. Athar \textit{et al.}, ``Status and Perspectives of Neutrino Physics," \href{https://www.sciencedirect.com/science/article/abs/pii/S0146641022000084?via%3Dihub}{Prog. Part. Nucl. Phys. \textbf{124}, 103947 (2022)} [\href{https://arxiv.org/abs/2111.07586}{arXiv:2111.07586}].

%\bibitem{DarkReview} L. Amendola {\it el al.}, ``Cosmology and fundamental physics with the Euclid satellite,'' \href{https://link.springer.com/article/10.1007/s41114-017-0010-3}{Living Rev. Relativ. {\bf21}, 2 (2018)}.
%%%%%%%%%%%%%%%%%%%%%%%%%%%%%%%%%%%%%%%%%%%%%%%%%%%%%%%%%%
%%%%%%%%%%%%%%%%%%%%%%%%%%%%%%%%%%%%%%%%%%%%%%%%%%%%
%\bibitem{MultiReview1} M. Ahlers and F. Halzen, ``Opening a new window onto the universe with IceCube", \href{https://www.sciencedirect.com/science/article/abs/pii/S0146641018300346}{Prog. Part. Nucl. Phys. {\bf102}, 73 (2018)} [\href{https://arxiv.org/abs/1805.11112}{arXiv:1805.11112}].
%%%%%%%%%%%%%%%%%%%%%%%%%%%%%%%%%%%%%%%%%%%%%%%%%%%%%%%
%%%%%%%%%%%%%%%%%% Pontecorvo %%%%%%%%%%%%%%%%%%%%%%%%%
%%%%%%%%%%%%%%%%%%%%%%%%%%%%%%%%%%%%%%%%%%%%%%%%%%%%%%%
\bibitem{Pontecorvo1} B. Pontecorvo, ``Mesonium and anti-mesonium", Sov. Phys. JETP {\bf6}, 429 (1957).

\bibitem{Pontecorvo2} B. Pontecorvo, ``Inverse beta processes and nonconservation of lepton charge", Sov. Phys. JETP {\bf7}, 172 (1958).
%%%%%%%%%%%%%%%%%%%%%%%%%%%%%%%%%%%%%%%%%%%%%%%%%
%%%%%%%%%%%%%%%%%%%%%%%%%%%%%%%%%%%%%%%%%%%%%%%%%
\bibitem{MSW1} L. Wolfenstein, ``Neutrino oscillations in matter,'' \href{https://journals.aps.org/prd/abstract/10.1103/PhysRevD.17.2369}{Phys. Rev. D{\bf17}, 2369 (1978)}.

\bibitem{MSW2} S.\,P. Mikheyev and A.\,Yu. Smirnov, ``Resonant amplification of $\nu$ oscillations in matter and solar-neutrino spectroscopy,'' \href{https://link.springer.com/article/10.1007/BF02508049}{Il Nuovo Cimento C{\bf9}, 17 (1986)}.
%%%%%%%%%%%%%%%%%%%%%%%%%%%%%%%%%%%%%%%%%%%%%%%%%
%%%%%%%%%%%%%%%%%%%%%%%%%%%%%%%%%%%%%%%%%%%%%%%%%
%%%%%%
%%%%%%%%%%% History of observations of NO %%%%%%%%%%%%%
%%%%%%%%%%%%%%%%%%%%%%%%%%%%%%%%%%%%%%%%%%%%%%%%%%%%%%%
%\bibitem{HistoryNO0} R. Davis, D.\,S. Harmer, and K.\,C. Hoffman. ``Search for Neutrinos from the Sun," \href{https://journals.aps.org/prl/abstract/10.1103/PhysRevLett.20.1205}{Phys. Rev. Lett. {\bf20}, 1205 (1968)}.

%\bibitem{HistoryNO1} B.\,T. Cleveland {\it et al}., ``Measurement of the Solar Electron Neutrino Flux with the Homestake Chlorine Detector,'' \href{https://iopscience.iop.org/article/10.1086/305343}{ApJ {\bf496}, 505 (1998)}.


%\bibitem{HistoryNO2} Y. Fukuda {\it et al}. (Super-Kamiokande Collaboration), ``Evidence for Oscillation of Atmospheric Neutrinos,'' \href{https://journals.aps.org/prl/abstract/10.1103/PhysRevLett.81.1562}{Phys. Rev. Lett. {\bf81}, 1562 (1998)} [\href{https://arxiv.org/abs/hep-ex/9807003}{arXiv:hep-ex/9807003}].

%\bibitem{HistoryNO3} Q.\,C. Ahmad {\it et al} (SNO Collaboration), ``Measurement of the Rate of
%$\nu_e + d \rightarrow p + p + e^-$ Interactions Produced by $\,^8$B Solar Neutrinos at the Sudbury Neutrino Observatory,'' \href{https://journals.aps.org/prl/abstract/10.1103/PhysRevLett.87.071301}{Phys. Rev. Lett. {\bf87}, 071301 (2001)} [\href{https://arxiv.org/abs/nucl-ex/0106015}{arXiv:nucl-ex/0106015}].


%\bibitem{HistoryNO4} M.\,H. Ahn {\it et al}. (K2K Collaboration), ``Indications of Neutrino Oscillation in a 250 km Long-baseline Experiment,'' \href{https://journals.aps.org/prl/abstract/10.1103/PhysRevLett.90.041801}{Phys. Rev. Lett. {\bf90}, 041801 (2003)} [\href{https://arxiv.org/abs/hep-ex/0212007v2}{arXiv:hep-ex/0212007}].

%\bibitem{HistoryNO5} S. Abe {\it et al}. (KamLAND Collaboration), ``Precision Measurement of Neutrino Oscillation Parameters with KamLAND,'' \href{https://journals.aps.org/prl/abstract/10.1103/PhysRevLett.100.221803}{Phys. Rev. Lett. {\bf100}, 221803 (2008)}; \href{https://journals.aps.org/prl/abstract/10.1103/PhysRevLett.101.119904}{Errata Phys. Rev. Lett. 101, 119904 (2008)}; \href{https://journals.aps.org/prl/abstract/10.1103/PhysRevLett.101.259901}{Phys. Rev. Lett. 101, 259901 (2008)} [\href{https://arxiv.org/abs/0801.4589}{arXiv:0801.4589}].

%\bibitem{HistoryNO6} J.\,N. Abdurashitov {\it et al}., ``Measurement of the solar neutrino capture rate with gallium metal. III: Results for the 2002--2007 data-taking period,'' \href{https://journals.aps.org/prc/abstract/10.1103/PhysRevC.80.015807}{Phys. Rev. C {\bf80}, 015807 (2009)} [\href{https://arxiv.org/abs/0901.2200}{arXiv:0901.2200}].

%\bibitem{HistoryNO7} F.\,P. An {\it et al}. (KamLAND Collaboration), ``Observation of Electron-Antineutrino Disappearance at Daya Bay,'' \href{https://journals.aps.org/prl/abstract/10.1103/PhysRevLett.108.171803}{Phys. Rev. Lett. {\bf108}, 171803 (2012)} [\href{https://arxiv.org/abs/1203.1669}{arXiv:1203.1669}].

%\bibitem{HistoryNO8} J.\,K. Ahn {\it et al}. (RENO Collaboration), ``Observation of Reactor Electron Antineutrinos Disappearance in the RENO Experiment,'' \href{https://journals.aps.org/prl/abstract/10.1103/PhysRevLett.108.191802}{Phys. Rev. Lett. 108, 191802 (2012)} [\href{https://arxiv.org/abs/1204.0626}{arXiv:1204.0626}].

%\bibitem{HistoryNO9} Y. Abe {\it et al}. (KamLAND Collaboration), ``Reactor electron antineutrino disappearance in the Double Chooz experiment,'' \href{https://journals.aps.org/prd/abstract/10.1103/PhysRevD.86.052008}{Phys. Rev. D {\bf86}, 052008 (2012)} [\href{https://arxiv.org/abs/1207.6632}{arXiv:1207.6632}].


%\bibitem{HistoryNO10} F. Kaether {\it et al}., ``Reanalysis of the GALLEX solar neutrino flux and source experiments,'' \href{https://www.sciencedirect.com/science/article/pii/S0370269310000729?via%3Dihub}{Phys. Lett. B {\bf685}, 47 (2010)} [\href{https://arxiv.org/abs/1001.2731}{arXiv:1001.2731}].


%\bibitem{HistoryNO11} P. Adamson, {\it et al}. (MINOS Collaboration), ``Combined analysis of $\nu_\mu$ disappearance and $\nu_\mu\rightarrow\nu_e$ appearance in MINOS using accelerator and atmospheric neutrinos,'' \href{https://journals.aps.org/prl/abstract/10.1103/PhysRevLett.112.191801}{Phys. Rev. Lett. {\bf112}, 191801 (2014)} [\href{https://arxiv.org/abs/1403.0867}{arXiv:1403.0867}].

\bibitem{SolarReview2022} X-J. Xu, Z. Wang and S. Chen, ``Solar neutrino physics,'' \href{https://www.sciencedirect.com/science/article/abs/pii/S0146641023000248?via%3Dihub}{Prog. Part. Nucl. Phys. \textbf{131}, 104043 (2023)} [\href{https://arxiv.org/abs/2209.14832}{arXiv:2209.14832}].


%%%%%%%%%%%%%%%%% List %%%%%%%%%%%%%%%%%
%%%%%%%%%%%%%%%%%%%%%%%%%%%%%%%%%%%%%%%%%%%%%%%%%%%%%%%
%%%%%%%%%%%%% Gravitational Effects %%%%%%%%%%%%%%%%%%%
%%%%%%%%%%%%%%%%%%%%%%%%%%%%%%%%%%%%%%%%%%%%%%%%%%%%%%%

%\bibitem{Gasperini} M. Gasperini, ``Experimental constraints on a minimal and nonminimal violation of the equivalence principle in the oscillations of massive neutrinos,'' \href{https://journals.aps.org/prd/abstract/10.1103/PhysRevD.39.3606}{Phys. Rev. D {\bf39}, 3606 (1989)}.

%\bibitem{Wudka} J. Wudka, ``Gravitational Effects on Neutrino Oscillations", \href{https://www.worldscientific.com/doi/abs/10.1142/S0217732391003808}{Mod. Phys. Lett. A {\bf06}(36), 3291 (1991)}.


%\bibitem{Ahluwalia1} D.\,V. Ahluwalia and C. Burgard, ``Gravitationally Induced Neutrino-Oscillation Phases,'' \href{https://link.springer.com/article/10.1007/BF03218936}{Gen. Rel. Grav. {\bf28}, 1161 (1996)} [\href{https://arxiv.org/abs/gr-qc/9603008v6}{arXiv:gr-qc/9603008}].

%\bibitem{Bhattacharya} T. Bhattacharya, S. Habib and E. Mottola, ``Gravitationally induced neutrino oscillation phases in static spacetimes,'' \href{https://journals.aps.org/prd/abstract/10.1103/PhysRevD.59.067301}{Phys. Rev. D {\bf59} 067301 (1999)} [\href{https://arxiv.org/abs/gr-qc/9605074}{arXiv:gr-qc/9605074}].

%\bibitem{Cardall} C.\,Y. Cardall and G.\,M. Fuller, ``Neutrino oscillations in curved spacetime: an heuristic treatment,'' \href{https://journals.aps.org/prd/abstract/10.1103/PhysRevD.55.7960}{Phys. Rev. D {\bf55}, 7960 (1997)} [\href{https://arxiv.org/abs/hep-ph/9610494v1}{arXiv:hep-ph/9610494}].

%\bibitem{Fornengo1} N. Fornengo {\it et al.}, ``Gravitational Effects on the Neutrino Oscillation,'' \href{https://journals.aps.org/prd/abstract/10.1103/PhysRevD.56.1895}{Phys. Rev. D {\bf56}, 1895 (1997)} [\href{https://arxiv.org/abs/hep-ph/9611231v2}{arXiv:hep-ph/9611231}].


%\bibitem{Fornengo2} N. Fornengo {\it et al.}, ``Gravitational effects on the neutrino oscillation in vacuum,'' \href{https://www.sciencedirect.com/science/article/abs/pii/S0920563298004356?via%3Dihub}{Nucl. Phys. {\bf B} - Proc. Suppl. {\bf70}, 264 (1999)} [\href{https://arxiv.org/abs/hep-ph/9711494v1}{arXiv:hep-ph/9711494}].

%\bibitem{Remarks} J.\,G. Pereira and C.\,M. Zhang, ``Some Remarks on the Neutrino Oscillation Phase in a Gravitational Field,'' \href{https://link.springer.com/article/10.1023%2FA%3A1001902706237}{Gen. Rel. Grav. {\bf32}, 1633 (2000)} [\href{https://arxiv.org/abs/gr-qc/0002066v3}{arXiv:gr-qc/0002066}].


%\bibitem{InsideNO1} C.\,M. Zhang and A. Beesham, ``The general treatment of high/low energy particle interference phase in a gravitational field,'' \href{https://link.springer.com/article/10.1023%2FA%3A1010224214296}{Gen. Rel. Grav. {\bf33}, 1011 (2001)} [\href{https://arxiv.org/abs/gr-qc/0004048v4}{arXiv:gr-qc/0004048}].


%\bibitem{NeutrinoInCS} R\,.M. Crocker, C. Giunti, and D.\,J. Mortlock, ``Neutrino Interferometry In Curved Spacetime,'' \href{https://journals.aps.org/prd/abstract/10.1103/PhysRevD.69.063008}{Phys. Rev. D {\bf69} 063008 (2004)} [\href{https://arxiv.org/abs/hep-ph/0308168v2}{arXiv:hep-ph/0308168v2}].

%\bibitem{Lambiase1} G. Lambiase {\it et al.}, ``Neutrino optics and oscillations in gravitational fields,'' \href{https://journals.aps.org/prd/abstract/10.1103/PhysRevD.71.073011}{Phys. Rev. D {\bf71}, 073011 (2005)} [\href{https://arxiv.org/abs/gr-qc/0503027v1}{arXiv:gr-qc/0503027}].

%\bibitem{HamiltonJacobi1} S.\,I. Godunov and G.\,S. Pastukhov, ``Neutrino Oscillations in Gravitational Field,'' \href{https://link.springer.com/article/10.1134%2FS1063778811020104}{Phys. Atom. Nucl. {\bf74}, 302 (2011)} [\href{https://arxiv.org/abs/0906.5556v2}{arXiv:0906.5556}].

%\bibitem{HamiltonJacobi2} L. Visinelli, ``Neutrino flavor oscillations in a curved space-time,'' \href{https://link.springer.com/article/10.1007%2Fs10714-015-1899-z}{Gen. Rel. Grav. {\bf47}, 62 (2015)} [\href{https://arxiv.org/abs/1410.1523}{arXiv:1410.1523}].


%\bibitem{InsideNO2} Yu-H. Zhang and Xue-Q. Li, ``Three-generation neutrino oscillations in curved spacetime,'' \href{https://www.sciencedirect.com/science/article/pii/S0550321316302565?via%3Dihub}{Nucl. Phys. B {\bf911}, 563 (2016)} [\href{https://arxiv.org/abs/1606.05960}{arXiv:1606.05960}].

%\bibitem{SuperNovaNO} Y. Yang and J.\,P. Kneller, ``GR Effects in Supernova Neutrino Flavor Transformation,'' \href{https://journals.aps.org/prd/abstract/10.1103/PhysRevD.96.023009}{Phys. Rev. D {\bf96}, 023009 (2017)} [\href{https://arxiv.org/abs/1705.09723v3}{arXiv:1705.09723}].


%\bibitem{Dvornikov} M. Dvornikov, ``Neutrino flavor oscillations in stochastic gravitational waves,'' \href{https://journals.aps.org/prd/abstract/10.1103/PhysRevD.100.096014}{Phys. Rev. D {\bf100}, 096014 (2019)} [\href{https://arxiv.org/abs/1906.06167}{arXiv:1906.06167}].

%\bibitem{SpinNO} M. Dvornikov, ``Neutrino spin oscillations in external fields in curved spacetime,'' \href{https://journals.aps.org/prd/abstract/10.1103/PhysRevD.99.116021}{Phys. Rev. D {\bf99}, 116021 (2019)} [\href{https://arxiv.org/abs/1902.11285}{arXiv:1902.11285}].

%\bibitem{Geometric} S. Chakrabarty and A. Lahiri, ``Geometrical contribution to neutrino mass matrix,'' \href{https://link.springer.com/article/10.1140/epjc/s10052-019-7209-2}{Eur. Phys. J. C {\bf79}, 697 (2019)} [\href{https://arxiv.org/abs/1904.06036}{arXiv:1904.06036}].

%\bibitem{Koutsoumbas} G. Koutsoumbas and D. Metaxas, ``Neutrino oscillations in gravitational and cosmological backgrounds,'' \href{https://link.springer.com/article/10.1007%2Fs10714-020-02758-z}{Gen. Relativ. Gravit. {\bf52}, 102 (2020)} [\href{https://arxiv.org/abs/1909.02735}{arXiv:1909.02735}].

%%%%%%%%%%%%%%%%%%%%%%%%%%%%%%%%%%%%%%%%%%%%%%%%%%%%%%%
%%%%%%%%%%% Equivalence Principle %%%%%%%%%%%%%%%%%%%%%
%%%%%%%%%%%%%%%%%%%%%%%%%%%%%%%%%%%%%%%%%%%%%%%%%%%%%%%

%\bibitem{Blasone3} M. Blasone {\it et al}., ``Non-relativistic neutrinos and the weak equivalence principle apparent violation,'' \href{https://www.sciencedirect.com/science/article/pii/S0370269320306869?via%3Dihub}{Phys. Lett. B {\bf811}, 135883 (2020)} [\href{https://arxiv.org/abs/2001.09974}{arXiv:2001.09974}].

%\bibitem{Capolupo} A. Capolupo, G. Lambiase and A. Quaranta, ``Neutrinos in curved spacetime: Particle mixing and flavor oscillations,'' \href{https://journals.aps.org/prd/abstract/10.1103/PhysRevD.101.095022}{Phys. Rev. D {\bf101}, 095022 (2020)} [\href{https://arxiv.org/abs/2003.00516}{arXiv:2003.00516}].






%%%%%%%%%%%%%%%%%%%%%%%%%%%%%%%%%%%%%%%%%%%%%%%%%%%
%%%%%%%%%%%%%%%% Cosmology %%%%%%%%%
%%%%%%%%%%%%%%%%%%%%%%%%%%%%%%%%%%%%%%%%%%%%%%%%%%%%%%%
%\bibitem{Xiu-Ju} H. Xiu-Ju, L. Ze-Jun and W. Yong-Jiu, ``Mass neutrino oscillations in Robertson–Walker space–time,'' \href{https://iopscience.iop.org/article/10.1088/1009-1963/15/1/038/meta}{Chinese Phys. {\bf15}, 229 (2006)}.

%\bibitem{Xin-Lian} L. Xin-Lian, B. Hua and L. Sheng-Peng, ``Cosmic Neutrino Oscillation or Separation,'' \href{https://iopscience.iop.org/article/10.1088/0256-307X/23/10/019/meta}{Chinese Phys. Lett. {\bf23}, 2691 (2006)}.

%\bibitem{Ren} J. Ren and H. Liu , ``Neutrino Oscillations in the Robertson-Walker Metric and the Cosmological Blue Shift of the Oscillation Length,'' \href{https://link.springer.com/article/10.1007/s10773-010-0473-4}{Int. J. Theor. Phys. {\bf49}, 2805 (2010)}

%\bibitem{JunPen} R. Jun and G. Jin-Peng, ``Interference Phase of Neutrino Oscillation in Schwarzschild-de Sitter Space-Time,'' \href{https://iopscience.iop.org/article/10.1088/0253-6102/53/4/16/meta}{Commun. Theor. Phys. {\bf53}, 665 (2010)}.

%\bibitem{Ren2} J. Ren and C-M. Zhang, ``Neutrino oscillations in Kerr-Newman space-time,'' \href{https://iopscience.iop.org/article/10.1088/0264-9381/27/6/065011}{Class. Quantum Grav. {\bf27}, 065011 (2010)} [\href{https://arxiv.org/abs/1002.0648}{arXiv:1002.0648}].


%\bibitem{RenPan} J. Ren and Y-Y. Pan, ``Neutrino Oscillations in the de Sitter and the Anti-de Sitter Space-Time,'' \href{https://link.springer.com/article/10.1007/s10773-011-0757-3}{Int. J. Theor. Phys. {\bf50}, 2614 (2011)}.


%\bibitem{Tao} C. Guang-Tao and W. Yong-Jiu, ``Interference Phase of Mass Neutrino in Schwarzschild de Sitter Field,'' \href{https://iopscience.iop.org/article/10.1088/0256-307X/28/2/029701/meta}{Chinese Phys. Lett. {\bf28}, 029701 (2011)}.


%\bibitem{ChatelainVolpe} A. Chatelain and M.\,C. Volpe, ``Neutrino decoherence in presence of strong gravitational fields'', \href{https://www.sciencedirect.com/science/article/pii/S037026931930872X?via%3Dihub}{Phys. Lett. B {\bf801}, 135150 (2020)} [\href{https://arxiv.org/abs/1906.12152}{arXiv:1906.12152}].

%\bibitem{Petruzziello} L. Petruzziello, ``Comment on ``Neutrino decoherence in presence of strong gravitational fields"'', \href{https://www.sciencedirect.com/science/article/pii/S0370269320305876?via%3Dihub}{Phys. Lett. B {\bf809}, 135784 (2020)} [\href{https://arxiv.org/abs/2009.06044}{arXiv:2009.06044}].

%%%%%%%%%%%%%%%%%% Alternative Gravity %%%%%%%%%%%%%%%

%\bibitem{Chakraborty1} S. Chakraborty, ``Constraining Alternative Gravity Theories Using The Solar Neutrino Problem,'' \href{https://iopscience.iop.org/article/10.1088/0264-9381/31/5/055005}{Class. Quantum Grav. {\bf31}, 055005 (2014)} [\href{https://arxiv.org/abs/1309.0693}{arXiv:1309.0693}].


%\bibitem{Chakraborty2} S. Chakraborty, ``Aspects of Neutrino Oscillation in Alternative Gravity Theories,'' \href{https://iopscience.iop.org/article/10.1088/1475-7516/2015/10/019}{JCAP {\bf10}, 019 (2015)} [\href{https://arxiv.org/abs/1506.02647}{arXiv:1506.02647}].

%\bibitem{Mandal} S. Mandal, ``Neutrino oscillations in cosmological spacetime,'' \href{https://www.sciencedirect.com/science/article/pii/S0550321321000353}{Nucl. Phys. B {\bf965}, 115338 (2021)}.


%%%%%%%%%%%%%%%%%%%%%%%%%%%%%%%%%%%%%%%%%%%%%%%%%%%
%%%%%%%%%%%%%%%% Quantum Gravity %%%%%%%%%
%%%%%%%%%%%%%%%%%%%%%%%%%%%%%%%%%%%%%%%%%%%%%%%%%%%%%%%
%\bibitem{Alexandre1} J. Alexandre {\it et al.}, ``Neutrino oscillations in a stochastic model for space-time foam,'' \href{https://journals.aps.org/prd/abstract/10.1103/PhysRevD.77.105001}{Phys. Rev. D {\bf77}, 105001 (2008)} [\href{https://arxiv.org/abs/0712.1779}{arXiv:0712.1779}].

%\bibitem{Alexandre2} J. Alexandre {\it et al.}, ``Neutrino oscillations in a Robertson-Walker Universe with space time foam,'' \href{https://journals.aps.org/prd/abstract/10.1103/PhysRevD.79.107701}{Phys. Rev. D {\bf79}, 107701 (2009)} [\href{https://arxiv.org/abs/0902.3386}{arXiv:0902.3386}].

%\bibitem{Marletto} C. Marletto, V. Vedral and D. Deutsch, ``Quantum-gravity effects could in principle be witnessed in neutrino-like oscillations,'' \href{https://iopscience.iop.org/article/10.1088/1367-2630/aad5d8}{New J. Phys. {\bf20}, 083011 (2018)} [\href{https://arxiv.org/abs/1804.02662}{arXiv:1804.02662}].




%%%%%%%%%%%%%%%%%%%%%%%%%%%%%%%%%%%%%%%%%%%%%%%%%%%%%%%
%%%%%%%%%%%%%%%%%%% With Torsion %%%%%%%%%%%%%%%%%%%%%%
%%%%%%%%%%%%%%%%%%%%%%%%%%%%%%%%%%%%%%%%%%%%%%%%%%%%%%%
%\bibitem{Torsion1} M. Adak, T. Dereli and L.\,H. Ryder, ``Neutrino Oscillations Induced by Space-Time Torsion,'' \href{https://iopscience.iop.org/article/10.1088/0264-9381/18/8/307}{Class. Quant. Grav. {\bf18}, 1503 (2001)} [\href{https://arxiv.org/abs/gr-qc/0103046v1}{arXiv:gr-qc/0103046}].

%\bibitem{Torsion2} A.\,A. Sousa, D.\,M. Oliveira and R.\,B. Pereira, ``Space-time Torsion and Neutrino Oscillations in Vacuum,'' [\href{https://arxiv.org/abs/0912.1317}{arXiv:0912.1317}].


%%%%%%%%%%%%%%%%%%%%%%%%%%%%%%%%%%%%%%%%%%%%%%%%%%%%%%%
%%%%%%%%%%%% Extended Theories of Gravity %%%%%%%%%%%%%
%%%%%%%%%%%%%%%%%%%%%%%%%%%%%%%%%%%%%%%%%%%%%%%%%%%%%%%
%\bibitem{Chakraborty} S. Chakraborty, ``Aspects of Neutrino Oscillation in Alternative Gravity Theories,'' \href{https://iopscience.iop.org/article/10.1088/1475-7516/2015/10/019}{JCAP {\bf10}, 019 (2015)} [\href{https://arxiv.org/abs/1506.02647}{arXiv:1506.02647}].

%\bibitem{Antonelli} V. Antonelli, L. Miramonti and M.\,D.\,C. Torri, ``Neutrino oscillations and Lorentz invariance violation in a Finslerian geometrical model,'' \href{https://link.springer.com/article/10.1140/epjc/s10052-018-6124-2}{Eur. Phys. J. C {\bf78}, 667 (2018)}. 

%\bibitem{Buoninfante} L. Buoninfante {\it et al.}, ``Neutrino oscillations in extended theories of gravity,'' \href{https://journals.aps.org/prd/abstract/10.1103/PhysRevD.101.024016}{Phys. Rev. D {\bf101}, 024016 (2020)} [\href{https://arxiv.org/abs/1906.03131}{arXiv:1906.03131}].

%\bibitem{Arxiv2022} H.\,Y Ahmadabadi and H. Mohseni Sadjadi, ``Non-standard neutrino interaction induced by conformal coupling,'' \href{https://arxiv.org/abs/2006.01636}{arXiv:2006.01636}.

%\bibitem{ShortReview} G.\,G. Luciano and L. Petruzziello, ``Testing gravity with neutrinos: from classical to quantum regime,'' \href{https://www.worldscientific.com/doi/abs/10.1142/S0218271820430026}{Int. J. Mod. Phys. D {\bf29}, 2043002 (2020)} [\href{https://arxiv.org/abs/2007.08664v1}{arXiv:2007.08664}].


%\bibitem{ConformalNO} H.\,M. Sadjadi and H.\,Y. Ahmadabadi, ``Damped Neutrino Oscillations in a Conformal Coupling Model,'' \href{https://journals.aps.org/prd/abstract/10.1103/PhysRevD.103.065012}{Phys. Rev. D {\bf103}, 065012 (2021)} [\href{https://arxiv.org/abs/2012.03633}{arXiv:2012.03633}].


%%%%%%%%%%%%%%%%%%%%%%%%%%%%%%%%%%
%%%%%%%%%%%%%%%%%%%%%%%%%%%%%%%%%%%%%%%%%%%%%%%%%%%%%%%%%%
%%%%%%%%%%%%%%%%%%%%%%%%%%%%%%%%%%%%%%%%%%%%%%%%%%%%%%%%%%
%%%%%% Conformally Coupled Neutrino oscillations %%
%%%%%%%%%%%%%%%%%%%%%%%%%%%%%%%%%%%%%%%%%%%%%%%%%%%%%%%%%%
%%%%%%%%%%%%%%%%%%%%%%%%%%%%%%%%%%%%%%%%%%%%%%%%%%%%%%%%%%

%\bibitem{DarkU2021} A.\,R. Khalifeh and R. Jimenez, ``Distinguishing Dark Energy models with neutrino oscillations,'' \href{https://www.sciencedirect.com/science/article/pii/S2212686421001278?via%3Dihub}{Phys. Dark Universe {\bf34}, 100897 (2021)} [\href{https://arxiv.org/abs/2105.07973}{arXiv:2105.07973}].

\bibitem{NWPG} P\,. Sadeghi {\it et al}., ``Wave packet treatment of neutrino flavor oscillations in various spacetimes,'' \href{https://link.springer.com/article/10.1007/s10714-021-02872-6}{Gen. Relativ. Gravit. {\bf53}, 98 (2021)} [\href{https://arxiv.org/abs/2111.01441}{arXiv:2111.01441}].

\bibitem{Swami1} H. Swami, K. Lochan and K.\,M. Patel, ``Aspects of gravitational decoherence in neutrino lensing,'' \href{https://journals.aps.org/prd/abstract/10.1103/PhysRevD.104.095007}{Phys. Rev. D \textbf{104}, 095007 (2021)}.


\bibitem{ConfNO} F. Hammad, P. Sadeghi and N. Fleury, ``Plane-wave and wavepacket neutrino flavor oscillations in vacuum in conformal coupling models,'' \href{https://journals.aps.org/prd/abstract/10.1103/PhysRevD.106.065019}{Phys. Rev. D \textbf{106}, 065019 (2022)} [\href{https://arxiv.org/abs/2209.03899}{arXiv:2209.03899}].

\bibitem{Swami2} H. Swami, ``Neutrino flavor oscillations in a rotating spacetime,'' \href{https://link.springer.com/article/10.1140/epjc/s10052-022-10902-z}{Eur. Phys. J. C \textbf{82}, 974 (2022)}.

\bibitem{SNOMatter} F. Hammad, P. Sadeghi and N. Fleury, ``Neutrino flavor oscillations inside matter in conformal coupling models,'' \href{https://journals.aps.org/prd/abstract/10.1103/PhysRevD.107.104015}{Phys. Rev. D \textbf{107}, 104015 (2023)} [\href{https://arxiv.org/abs/2304.03746}{arXiv:2304.03746}].


%%%%%%%%%%%%%%%%%%%%%%%%%%%%%%%%%%%%%%%%%%%%%%%%%%%%%%%%%%
%%%%%%%%%%%%%%%%%%%%%%%%%%%%%%%%%%%%%%%%%%%%%%%%%%%%%%%%%% Early Proposals for Spin Flip
%%%%%%%%%%%%%%%%%%%%%%%%%%%%%%%%%%%%%%%%%%%%%%%%%%%%%%%%%%
\bibitem{Cisneros1970} A. Cisneros, ``Effect of Neutrino Magnetic Moment on Solar Neutrino Observations,'' \href{https://link.springer.com/article/10.1007/BF00654607}{Astrophys Space Sci. \textbf{10}, 87 (1971)}.

\bibitem{Lim} C-S. Lim and W.\,J. Marciano, ``Resonant spin-flavor precession of solar and supernova neutrinos,'' \href{https://journals.aps.org/prd/abstract/10.1103/PhysRevD.37.1368}{Phys. Rev. D \textbf{37}, 1368 (1988)}.


\bibitem{Akhmedov} E.\,Kh. Akhmedov, ``Resonant amplification of neutrino spin rotation in matter and the solar-neutrino problem,'' \href{https://www.sciencedirect.com/science/article/abs/pii/0370269388910489}{Phys. Lett. B \textbf{213}, 64 (1988)}.



%\bibitem{Fujikawa} K. Fujikawa and R.\,E. Shrock, ``Magnetic Moment of a Massive Neutrino and Neutrino-Spin Rotation,'' \href{https://journals.aps.org/prl/abstract/10.1103/PhysRevLett.45.963}{Phys. Rev. Lett. {\bf45}, 963 (1980)}.

%\bibitem{Schechter} J. Schechter and J.\,W. F. Valle, ``Majorana neutrinos and magnetic fields,'' \href{https://journals.aps.org/prd/abstract/10.1103/PhysRevD.24.1883}{Phys. Rev. D \textbf{24}, 1883 (1981)} \href{https://journals.aps.org/prd/abstract/10.1103/PhysRevD.25.283};{Erratum: Phys. Rev. D \textbf{25}, 283 (1982)}.

%\bibitem{Voloshin1} M.\,B. Voloshin and M.\,I. Vysotsky, ``Majorana neutrinos and magnetic fields,'' \href{https://journals.aps.org/prd/abstract/10.1103/PhysRevD.24.1883}{Phys. Rev. D \textbf{24}, 1883 (1981)} 

%\bibitem{Voloshin1986} A. M\,.B. Voloshin, M.\,I. Vysotskii and L.\,B. Okun, ``Neutrino Electrodynamics and Possible Consequences for Solar Neutrinos. Pages 344-350,'' in J.\,N. Bahcall, D. Jr. Raymond, P. Parker, A. Smirnov and R. Ulrich (eds) \textit{Solar Neutrinos: The First Thirty Years} (Taylor \& Francis Group, CRC Press, 2018).

%\bibitem{Lim1987} C-S. Lim and W.\,J. Marciano, ``Resonant spin-flavor precession of solar and supernova neutrinos. Pages 351-356,'' in J.\,N. Bahcall, D. Jr. Raymond, P. Parker, A. Smirnov and R. Ulrich (eds) \textit{Solar Neutrinos: The First Thirty Years} (Taylor \& Francis Group, CRC Press, 2018).

%\bibitem{Akhmedov1988} E.\,Kh. Akhmedov, ``Resonant Amplification of Neutrino Spin in Matter and the Solar-Neutrino Problem. Pages 357-361,'' in J.\,N. Bahcall, D. Jr. Raymond, P. Parker, A. Smirnov and R. Ulrich (eds) \textit{Solar Neutrinos: The First Thirty Years} (Taylor \& Francis Group, CRC Press, 2018).

\bibitem{Studenikin} A.\,I. Studenikin, ``Neutrinos in Electromagnetic Fields and Moving Media,'' \href{https://link.springer.com/article/10.1134/1.1755390}{Phys. Atom. Nuclei {\bf67}, 993 (2004)}.

\bibitem{Book1994} J.\,N. Bahcall, D.\,Jr. Raymond, P. Parker, A. Smirnov and R. Ulrich (eds) \textit{Solar Neutrinos: The First Thirty Years} (Taylor \& Francis Group, CRC Press, Boca Raton, 2018).

\bibitem{Chukhnova2021} A.\,V. Chukhnova and A.\,E. Lobanov, ``Resonance enhancement of neutrino oscillations due to transition magnetic moments,'' \href{https://link.springer.com/article/10.1140/epjc/s10052-021-09611-w}{Eur. Phys. J. C \textbf{81}, 821 (2021)}.


%%%%%%%%%%%%%%%%%%%%%%%%%%%%%%%%%%%%%%%%%%%%%%%%%%%%%%%%%
%%%%%%%%%%%%%%%%%%%%%%%%%%%%%%%%%%%%%%%%%%% Spin Gravity
%%%%%%%%%%%%%%
%%%%%%%%%%%%%%%%%%%%%%%%%%%%%%%%%%%%%%%%%
\bibitem{Papini1991} Y.\,Q. Cai and G. Papini, ``Neutrino helicity flip from gravity-spin coupling,'' \href{https://journals.aps.org/prl/abstract/10.1103/PhysRevLett.66.1259}{Phys. Rev. Lett. \textbf{66}, 1259 (1991)}.

\bibitem{Casini1994} H. Casini and R. Montemayor,``Chirality transitions in gravitational fields,'' \href{https://journals.aps.org/prd/abstract/10.1103/PhysRevD.50.7425}{Phys. Rev. D {\bf50}, 7425 (1994)}.

\bibitem{Papini2004} D. Singh, N. Mobed and G. Papini, ``Helicity Precession of Spin-1/2 Particles in Weak Inertial and Gravitational Fields,'' \href{https://iopscience.iop.org/article/10.1088/0305-4470/37/34/010}{J. Phys. A \textbf{37}, 8329 (2004)} [\href{https://arxiv.org/abs/hep-ph/0405296v1}{arXiv:hep-ph/0405296v1}]

\bibitem{Dolan2006} S. Dolan, C. Doran and A. Lasenby, ``Fermion scattering by a Schwarzschild black hole,'' \href{https://iopscience.iop.org/article/10.1088/0305-4470/37/34/010}{Phys. Rev. D\textbf{74}, 064005 (2006)} [\href{https://arxiv.org/abs/gr-qc/0605031}{arXiv:gr-qc/0605031}].

%%%%%%%%%%%%%%%%%%%%%%%%%%%%%%%%%%%%%%%%%%%%%%%%%
%%%%%%%%%%%%%%%%%%%%%%%%%%%%%%%%%%%%%%%%%%%%%%%%%
%%%%%%%%%%%  Recent Spin Oscillation %%%%%%%%%%%%%%%%
%%%%%%%%%%%%%%%%%%%%%%%%%%%%%%%%%%%%%%%%%%%%%%%%%
%%%%%%%%%%%%%%%%%%%%%%%%%%%%%%%%%%%%%%%%%%%%%%%%%
%%%%%%%%%%%%%%%%%%%%%%%%%%%%%%%%%%%%%%%%%%%%%%%%%%%%%%%%%%
\bibitem{Dvornikov2006} M. Dvornikov, ``Neutrino spin oscillations in gravitational fields,'' \href{https://www.worldscientific.com/doi/abs/10.1142/S021827180600870X}{Int. J. Mod. Phys. D \textbf{15}, 1017 (2006)} [\href{https://arxiv.org/abs/hep-ph/0601095}{hep-ph/0601095}].

\bibitem{Dvornikov2013} M. Dvornikov, ``Neutrino spin oscillations in matter under the influence of gravitational
and electromagnetic fields,'' \href{https://iopscience.iop.org/article/10.1088/1475-7516/2013/06/015}{JCAP \textbf{06}, 015 (2013)} [\href{https://arxiv.org/abs/1306.2659}{arXiv:1306.2659}].

\bibitem{Alavi} S. A. Alavi and S. Nodeh, ``Neutrino spin oscillations in gravitational fields in noncommutative spaces,'' \href{https://iopscience.iop.org/article/10.1088/0031-8949/90/3/035301}{Phys. Scripta \textbf{90}, 035301 (2015)} [\href{https://arxiv.org/abs/1301.5977}{arXiv:1301.5977}].

\bibitem{Dvornikov2019} M. Dvornikov, ``Neutrino spin oscillations in external fields in curved spacetime,'' \href{https://journals.aps.org/prd/abstract/10.1103/PhysRevD.99.116021}{Phys. Rev. D \textbf{99}, 116021 (2019)} [\href{https://arxiv.org/abs/1902.11285}{arXiv:1902.11285}].

\bibitem{Dvornikov2020a} M. Dvornikov, ``Spin effects in neutrino gravitational scattering,'' \href{https://journals.aps.org/prd/abstract/10.1103/PhysRevD.101.056018}{Phys. Rev. D \textbf{101}, 056018 (2020)} [\href{https://arxiv.org/abs/1911.08317}{arXiv:1911.08317}].

\bibitem{Dvornikov2020b} M. Dvornikov, ``Spin oscillations of neutrinos scattered off a rotating black hole,'' \href{https://link.springer.com/article/10.1140/epjc/s10052-020-8046-z}{Eur. Phys. J. C \textbf{80}, 474 (2020)}.

\bibitem{PRD2021} L. Mastrototaro and G. Lambiase, ``Neutrino spin oscillations in conformally gravity coupling models and quintessence surrounding a black hole,'' \href{https://journals.aps.org/prd/abstract/10.1103/PhysRevD.104.024021}{Phys. Rev. D {\bf104}, 024021 (2021)} [\href{https://arxiv.org/abs/2106.07665}{arXiv:2106.07665}].


%%%%%%%%%%%%%%%%%%%%%%%%%%%%%%%%%%%%%%%%%%%%%%%%%%%%%%%%%
%%%%%%%%%%%%%%%%%%%%%%%%%%%%%%%%%%%%%%%%%%%%%%%%%%%%%%%%%
%%%%%%%%%%%%%%%%%%%%%%%%%%%%%%%%%%%%%%%%%%%%%%%%%%%%%%%%%
%%%%%%%%%%%%%%%%%%%%%%%%%%%%%%%%%%%%%%%%%%%%%%%%%%%%%%%%%
%\bibitem{NBook} C. Giunti and C.\,W. Kim, {\it Fundamentals of Neutrino Physics and Astrophysics} (Oxford University Press, Oxford, 2007).

%%%%%%%%%%%%%%%%%%%%%%%%%%%%%%%%%%%%%%%%%%%%%%%%%%%%%%%%%
%%%%%%%%%%%%%%%%%%%%%%%%%%%%%%%%%%%%%%%%%%%%%%%%%%%%%%%%%
%%%%%%%%%%%%%%%%%%%%%%%%%%%%%%%%%%%%%%%%%%%%%%%%%%%%%%%%%
%%%%%%%%%%%%%%%%%%%%%%%%%%%%%%%%%%%%%%%%%%%%%%%%%%%%%%%%%
\bibitem{DeSitter} W. De Sitter, ``On Einstein's Theory of Gravitation and its Astronomical Consequences. Second Paper,'' \href{https://academic.oup.com/mnras/article/77/2/155/979347?login=false}{Mon. Not. R. Astron. Soc. \textbf{77}, 155 (1916)}.


\bibitem{LensThirring} J. Lense and H. Thirring, ``\"Uber die Einflu\ss  der Eigenrotation der Zentralk\"orper auf die Bewegung der Planeten und Monde nach der Einsteinschen Gravitationstheorie,'' Zeit. Phys. \textbf{19}, 156 (1918).

%%%%%%%% Dark Review %%%%%%%%%%%%
\bibitem{DarkReview} N. Frusciante and L. Perenon, ``Effective Field Theory of Dark Energy: a Review,'' \href{https://www.sciencedirect.com/science/article/abs/pii/S0370157320300375?via%3Dihub}{Phys. Rept. \textbf{857}, 1 (2020)} [\href{https://arxiv.org/abs/1907.03150}{arXiv:1907.03150}].

\bibitem{Chameleon} J. Khoury and A. Weltman, ``Chameleon Fields: Awaiting Surprises for Tests of Gravity in Space,'' \href{https://journals.aps.org/prl/abstract/10.1103/PhysRevLett.93.171104}{Phys. Rev. Lett. {\bf93}, 171104(2004)} [\href{https://arxiv.org/abs/astro-ph/0309300}{arXiv:astro-ph/0309300}]. 

%%%%%%%%%%%%%%%%%%%%%%%%%%%%%%%%%%%%%%
%%%%%%%%%%%%%%%%%%%%%%%%%%%%%%%%%%%%%%
%%%%%%%%%%%%%%%%%%%%%%%%%%%%%%%%%%%%%%
\bibitem{ChameleonIntroduction} T.\,P. Waterhouse, ``An Introduction to Chameleon Gravity,'' \href{https://arxiv.org/abs/astro-ph/0611816v1}{arXiv:astro-ph/0611816v1}.


%\bibitem{PRD2004} J. Khoury and A. Weltman, ``Chameleon cosmology,'' \href{https://journals.aps.org/prd/abstract/10.1103/PhysRevD.69.044026}{Phys. Rev. D {\bf69}, 044026 (2004)} [\href{https://arxiv.org/abs/astro-ph/0309411}{arXiv:astro-ph/0309411}].

%\bibitem{JCAP2009} S. Tsujikawa, T. Tamaki, and R. Tavakol, ``Chameleon scalar fields in relativistic gravitational backgrounds,'' \href{https://iopscience.iop.org/article/10.1088/1475-7516/2009/05/020}{J. Cosmol. Astropart. Phys. {\bf05}, 020 (2009)} [\href{https://arxiv.org/abs/0901.3226}{arXiv:0901.3226}].

%\bibitem{LRR2018} C. Burrage and J. Sakstein, ``Tests of chameleon gravity'' \href{https://link.springer.com/article/10.1007/s41114-018-0011-x}{Living Rev. Relativ. {\bf21}, 1 (2018)} [\href{https://arxiv.org/abs/1709.09071}{arXiv:1709.09071}].
%%%%%%%%%%%%%%%%%%%%%%%%%%%%%%%%%%%%%%%%%%%%%%%%%%%%%%%%%%
%%%%%%%%%%%%%%%%%%%%%%%%%%%%%%%%%%%%%%%%%%%%%%%%%%%%%%%%%%
%%%%%%%%%%%%%%%%%%%%%%%%%%%%%%%%%%%%%%%%%%%%%%%%%%%%%%%%%%
%%%%%%%%%%%%%%%%%%%%%%%%%%%%%%%%%%%%%%%%%%%%%%%%%%%%%%%%%%

%%%%%%%%%%%%%%%%%%%%%%%%%%%%%%%%%%%%%%%%%%%%%%%%%%%%%%%%%%
%%%%%%%%%%%%%%%%%%%%%%%%%%%%%%%%%%%%%%%%%%%%%%%%%%%%%%%%%%
%%%%%%%%%%%%%%%%%%%%%%%%%%%%%%%%%%%%%%%%%%%%%%%%%%%%%%%%%%
%%%% Symmetron %%%%%%%%%%%%%%%%%%%%%%%%%%%%%%%%%%%%%%%%%%%
%%%%%%%%%%%%%%%%%%%%%%%%%%%%%%%%%%%%%%%%%%%%%%%%%%%%%%%%%%
%%%%%%%%%%%%%%%%%%%%%%%%%%%%%%%%%%%%%%%%%%%%%%%%%%%%%%%%%%
%%%%%%%%%%%%%%%%%%%%%%%%%%%%%%%%%%%%%%%%%%%%%%%%%%%%%%%%%
\bibitem{Symmetron} K. Hinterbichler and J. Khoury, ``Symmetron Fields: Screening Long-Range Forces Through Local Symmetry Restoration,'' \href{https://journals.aps.org/prl/abstract/10.1103/PhysRevLett.104.231301}{Phys. Rev. Lett. {\bf104}, 231301 (2010)} [\href{https://arxiv.org/abs/1001.4525}{arXiv:1001.4525}].


%\bibitem{IntroVainshtein} E. Babichev and C. Deffayet, ``An introduction to the Vainshtein mechanism,'' \href{https://iopscience.iop.org/article/10.1088/0264-9381/30/18/184001}{	Class. Quantum Grav. {\bf30}, 184001 (2013)} [\href{https://arxiv.org/abs/1304.7240}{arXiv:1304.7240}].

%\bibitem{Dilaton} P. Brax {\it el al}., ``The Dilaton and Modified Gravity, \href{}{Phys. Rev. D {\bf 82}, 063519 (2010)} [\href{https://arxiv.org/abs/1005.3735}{arXiv:1005.3735}].

\bibitem{ChameleonTests} C. Burrage and J. Sakstein, ``Tests of Chameleon Gravity,'' \href{https://link.springer.com/article/10.1007/s41114-018-0011-x}{Living Rev. Relativity {\bf21}, 1 (2018)} [\href{https://arxiv.org/abs/1709.09071}{arXiv:1709.09071}].

%%%%%%%%%%%%%%%%%%%%%%%%%%%%%%

%%%%%% MDP Equations %%%%%%%%%%%
\bibitem{Mathisson} M. Mathisson, ``Neue Mechanik materieller Systeme,'' \href{https://sbc.org.pl/dlibra/publication/323471/edition/305634/content}{Acta Phys. Polon. {\bf6}, 163 (1937)} [Translated and republished in \href{https://link.springer.com/article/10.1007/s10714-010-0939-y}{Gen. Relativ. Gravit. \textbf{42}, 1011 (2010)}].

\bibitem{Papapetrou} A. Papapetrou, ``Spinning Test-Particles in General Relativity. I.,'' \href{https://royalsocietypublishing.org/doi/10.1098/rspa.1951.0200}{Proc. R. Soc. London A {\bf209}, 248 (1951)}.

%\bibitem{Tulczyjew} W.\,M. Tulczyjew, ``Motion of multipole particles in general relativity theory binaries,'' \href{}{Acta Phys. Polon. 18, 393 (1959)}.

\bibitem{Dixon} W.\,G. Dixon, ``Extended Bodies in General Relativity: Their Description and Motion. Pages 156-219,'' in J. Ehlers (ed), \textit{Isolated Gravitating Systems in General Relativity} (Amsterdam, North-Holland, 1979).

\bibitem{BakerReview} B.\,M. Baker and R.\,F. O'Connell, ``The Gravitational Interaction: Spin, Rotation, and Quantum Effects--A Review,'' \href{https://link.springer.com/article/10.1007/BF00756587}{Gen. Relativ. Gravit. {\bf11}, 149 (1979)}.

\bibitem{Deriglazov} A.\,A. Deriglazov and W. Guzm\'an Ram\'irez, ``Recent progress on the description of relativistic spin: vector model of spinning particle and rotating body with gravimagnetic moment in General Relativity,'' \href{https://www.hindawi.com/journals/amp/2017/7397159/}{Advances in Mathematical Physics \textbf{2017}, 49 (2017)} [\href{https://arxiv.org/abs/1710.07135}{arXiv:1710.07135}].

%%%%%%%%%%%%%%%%%%%%%%%%%%%%%%%

\bibitem{Rudiger} R. R\"udiger, ``The Dirac Equation and Spinning Particles in General Relativity,'' \href{https://iopscience.iop.org/article/10.1088/0305-4470/14/2/017}{Proc. R. Soc. Lond. A \textbf{377}, 417 (1981)}.


\bibitem{Audretsch} J. Audretsch, ``Trajectories and spin motion of massive spin-1/2 particles in gravitational fields,'' \href{https://iopscience.iop.org/article/10.1088/0305-4470/14/2/017}{J. Phys. A: Math. Gen. \textbf{14}, 411 (1981)}.

\bibitem{CianfraniMontani1} F. Cianfrani and G. Montani, ``Curvature-spin coupling from the semi-classical limit of the Dirac equation,'' \href{https://www.worldscientific.com/doi/abs/10.1142/S0217751X08040214}{Int. J. Mod. Phys. A \textbf{23}, 1274 (2008)} [\href{https://arxiv.org/abs/0805.2480}{arXiv:0805.2480}].

\bibitem{CianfraniMontani2} F. Cianfrani and G. Montani, ``Dirac equations in curved space-time vs. Papapetrou spinning particles,'' \href{https://iopscience.iop.org/article/10.1209/0295-5075/84/30008}{EPL \textbf{84}, 30008 (2008)} [\href{https://arxiv.org/abs/0810.0447}{arXiv:0810.0447}].


%%%%%%%%%%%%%%%%%%%%%%%%%%%%%%%
%%%%%%%%%%%%%%%%%%%%%%%%%%%%%%%
%%%%%%%%%%%%%%%%%%%%%%%%%%%%%%%
%%%%%%%%%%%%%%%%%%%%%%%%%%%%%%%

\bibitem{2ndMethod} Y.\,N. Obukhov, A.\,J. Silenko and O.\,V. Teryaev, ``General treatment of quantum and classical spinning particles in external fields,'' \href{https://journals.aps.org/prd/abstract/10.1103/PhysRevD.96.105005}{Phys. Rev. D 96, 105005 (2017)} [\href{https://arxiv.org/abs/1708.05601}{arXiv:1708.05601}].

\bibitem{Oancea} M.\,A. Oancea and A. Kumar, ``Semiclassical analysis of Dirac fields on curved spacetime,'' \href{https://journals.aps.org/prd/abstract/10.1103/PhysRevD.107.044029}{Phys. Rev. D \textbf{107}, 044029 (2023)} [\href{https://arxiv.org/abs/2212.04414}{arXiv:2212.04414}].

%%%%%%%%%%%%%%%%%%%%%%%%%%%%%%%%%%%%%%%%%%%%%%%%%%%%%%%%%
%%%%%%%%%%%%%%%%%%%%%%%%%%%%%%%%%%%%%%%%%%%%%%%%%%%%%%%%%
\bibitem{BMT} V. Bargmann, L. Michel and V.\,L. Telegdi, ``Precession of the Polarization of Particles moving in a Homogeneous electromagnetic Field,'' \href{https://journals.aps.org/prl/abstract/10.1103/PhysRevLett.2.435}{Phys. Rev. Lett. {\bf2}, 435 (1959)}.

\bibitem{Mashhoon} B. Mashhoon, ``Particles with Spin in a Gravitational Field,'' \href{https://aip.scitation.org/doi/abs/10.1063/1.1665699}{J. Math. Phys. {\bf12}, 1075 (1971)}.
%%%%%%%%%%%%%%%%%%%%%%%%%%%%%%%%%
%%%%%%%%%%%%%%%%%%%%%%%%%%%%%%%%%


\bibitem{MTW} C.\,W. Misner, K.\,S. Thorne and J.\,A. Wheeler, \textit{Gravitation} (W.\,H. Freeman and Company, San Francisco, 1973).

%%%%%%%%%%%%%%%%%%%%%%%%%%%%%%%%%%%%%%%%%%%%%%%%%%%%%%%%%
%%%%%%%%% MDP equations %%%%%%%%%%%%

%%%%%%%%%%%%%%%%%%%%%%%%%%%%%%%%%%%%%%%%%%%%%%%%%%%%%%%%%
%%%%%%%%%%%%%%%%%%%%%%%%%%%%%%%%%%%%%%%%%%%%%%%%%%%%%%%%%
%%%%%%%%%%%%%%%%%%%%%%%%%%%%%%%%%%%%%%%%%

\bibitem{Pomeranskii1997} I.\,B. Khriplovich, A.\,A. Pomeransky, ``Equations of Motion of Spinning Relativistic Particle in External Fields,'' \href{https://link.springer.com/article/10.1134/1.558554}{J. Exp. Theor. Phys. \textbf{86}, 839 (1998)} [\href{https://arxiv.org/abs/gr-qc/9710098v1}{arXiv:gr-qc/9710098v1}].

\bibitem{Pomeranskii2000} A.\,A. Pomeranskii, R.\,A.Sen'kov and I.\,B. Khriplovich, ``Spinning Relativistic particles in external fields,'' \href{https://iopscience.iop.org/article/10.1070/PU2000v043n10ABEH000674/meta}{Phys.-Usp. \textbf{43}, 1055 (2000)}.

\bibitem{QED} V.\,B. Berestetskii, E.\,M. Lifshitz and L.\,P. Pitaevskii, \textit{Quantum Electrodynamics} (Pergamon Press, Oxford, 1982).

%%%%%%%%%%%%%%%%%%%%%%%
%%%%%%%%%%%%%%%%%%%%%%%%%%%%%%%%%%%%%%%%%%%%%%%%%%%%%%%%%%
%%%%%%%%%%%%%%%%%%%%%%%%%%%%%%%%%%%%%%%%%%%%%%%%%%%%%%%%%%

\bibitem{Wald} R.\,M. Wald, {\it General Relativity} (University of Chicago Press, Chicago, 1984).

\bibitem{Damour1} T. Damour, G.\,W. Gibbons and C. Gundlach, ``Dark matter, time-varying G, and a dilaton field,'' \href{https://journals.aps.org/prl/abstract/10.1103/PhysRevLett.64.123}{Phys. Rev. Lett. \textbf{64}, 123 (1990)}.

\bibitem{Damour2} T. Damour and A.\,M. Polyakov, ``The string dilation and a least coupling principle,'' \href{https://www.sciencedirect.com/science/article/abs/pii/0550321394901430}{Nucl. Phys. B \textbf{423}, 532 (1994)}.

\bibitem{SchroDi} N. Fleury, F. Hammad and  P. Sadeghi, ``Revisiting the Schrödinger-Dirac equation,'' \href{https://doi.org/10.3390/sym15020432}{Symmetry {\bf15}(2), 432 (2023)} [\href{https://arxiv.org/abs/2302.06723}{arXiv:2302.06723}].

\bibitem{Gravdi} F. Hammad, A. Landry and P. Sadeghi, ``Spin-1/2 particles under the influence of a uniform magnetic field in the interior Schwarzschild solution,'' \href{https://www.mdpi.com/2218-1997/7/12/467}{Universe \textbf{7}(12), 467 (2021)} [\href{https://arxiv.org/abs/2111.15448}{arXiv:2111.15448}].

%\bibitem{QFT2} N.\,D. Birrell and P.\,C.\,W. Davies, \textit{Quantum Fields in Curved Space} (Cambridge University Press, Cambridge, 1994). 


%%%%%%%%%%%%%%%%
%%%%%%%%%%%%%%%%%%%%%%%%%%%%%%%%%%%%%%%%%%%%%%%%%%%%%%%%%
%%%%%%%%%%%%%%%%%%%%%%%%%%%%%%%%%%%%%%%%%%%%%%%%%%%%%%%%%
%%%%%%%%%%%%%%%%%%%%%%%%%%%%%%%%%%%%%%%%%%%%%%%%%%%%%%%%%
\bibitem{Hartle} J.\,B. Hartle, \textit{Gravity: An Introduction to Einstein's General Relativity} (Pearson, Essex, 2014).



%%%%%%%%%%%%%%%%%%%%%%%%%%%%%%%%%%%%%%%%%%%%%%%%%%%%%%%
%%%%%%%%%%%%% Wave packet 1 %%%%%%%%%%%%%%%%%%%%%%%%%%%%%
%%%%%%%%%%%%%%%%%%%%%%%%%%%%%%%%%%%%%%%%%%%%%%%%%%%%%%%
%%%%%%%%%%%%%%%%%%%%%%%%%%%%%%%%%%%%%%%%%%%%%%%%%%%%%%%%%%
%%%% Chameleon %%%%%%%%%%%%%%%%%%%%%%%%%%%%%%%%%%%%%%%%%%%
%%%%%%%%%%%%%%%%%%%%%%%%%%%%%%%%%%%%%%%%%%%%%%%%%%%%%%%%%%
%%%%%%%%%%%%%%%%%%%%%%%%%%%%%%%%%%%%%%%%%%%%%%%%%%%%%%%%%%
%%%%%%%%%%%%%%%%%%%%%%%%%%%%%%%%%%%%%%%%%%%%%%%%%%%%%%%%%



%\bibitem{PRD2011} K. Hinterbichler, J. Khoury, A. Levy, and A. Matas, ``Symmetron cosmology,'' \href{https://journals.aps.org/prd/abstract/10.1103/PhysRevD.84.103521}{Phys.
%Rev. D {\bf84}, 103521 (2011)} [\href{https://arxiv.org/abs/1107.2112}{arXiv:1107.2112}].
%%%%%%%%%%%%%%%%%%%%%%%%%%%%%%%%%%%%%%%%%%%%%%%%%%%%%%%%%%
%%%%%%%%%%%%%%%%%%%%%%%%%%%%%%%%%%


%\bibitem{Will2018} C.\,M. Will, {\it Theory and Experiment in Gravitational Physics}, 2nd Edition (Cambridge University Press, Cambridge, 2018).
%%%%%%%%%%%%%%%%%%%%%%%%%%%%%%%%%%%%%%%%%%%%%%%%%
%%%%%%%%%%%%%%%%%%%%%%%%%%%%%%%%%%%%%%%%%%%%%%%%%
%%%%%%%%%%%%%%%%%%%%%%%%%%%%%%%%%%%%%%%%%%%%%%%%%%%%%%%%%
%%%%%%%%%%%%%%%%%%%%%%%%%%%%%%%%%%%%%%%%%%%%%%%%%%%%%%%%%
%%%%%%%%%%%%%%%%%%%%%%%%%%%%%%%%%%%%%%%%%%%%%%%%%
%%%%%%%%%%%%%%%%%%%%%%%%%%%%%%%%%%%%%%%%%%%%%%%%%
%%%%%%%%%%%%%%%%%%%%%%%%%%%%%%%%%%%%%%%%%%%%%%%%%%%%%%%

%%%%%%%%%%%%%%%%%%%%%%%%%%%%%%%%%%%%%%%%%%%%%%%%%%%%%%%%%%
%%%%%%%%%%%%%%%%%%%%%%%%%%%%%%%%%%%%%%%%%%%%%%%%%%%%%%%%%%
%%%%%%%%%%%%%%%%%%%%%%%%%%%%%%%%%%%%%%%%%%%%%%%%%%%%%%%%%%
%%%%%%%%%%%%%%%%%%%%%%%%%%%%%%%%%%%%%%%%%%%%%%%%%%%%%%%%%%

%%%%%%%%%%%%%%%%%%%%%%%%%%%%%%%%%%%%%%%%%%%%%%%%%%%%%%%%%%
%%%%%%%%%%%%%%%%%%%%%%%%%%%%%%%%%%%%%%%%%%%%%%%%%%%%%%%%%%
%%%% Mass-Varying Solar Neutrinos and their oscillations %
%%%%%%%%%%%%%%%%%%%%%%%%%%%%%%%%%%%%%%%%%%%%%%%%%%%%%%%%%%
%%%%%%%%%%%%%%%%%%%%%%%%%%%%%%%%%%%%%%%%%%%%%%%%%%%%%%%%%%
%%%%%%%%%%%%%%%%%%%%%%%%%%%%%%%%%%%%%%%%%%%%%%%%%%%%%%%%%
%%%%%%%%%%%
%%%%%%%%%%%%%%%%%%%%%%%%%%%%%%%%%%%%%%%%%%%%%%%%%%%%%%%%%%
%%%%%%%%%%%%%%%%%%%%%%%%%%%%%%%%%%%%%%%%%%%%%%%%%%%%%%%%%%
%%%%%%%%%%%%%%%%%%%%%%%%%%%%%%%%%%%%%%%%%%%%%%%%%%%%%%%%%%
%%%%%%%%%%%%%%%%%%%%%%%%%%%%%%%%%%%%%%%%%%%%%%%%%%%%%%%%%%
%%%%%%%%%%%%%%%%%%%%%%%%%%%%%%%%%%%%%%%%%%%%%%%%%
%%%%%%%%%%%%%%%%%%%%%%%%%%%%%%%%%%%%%%%%%%%%%%%%%
%%%%%%%%%%%%%%%%%%%%%%%%%%%%%%%%%%%%%%%%%%%%%%%%%
%%%%%%%%%%%%%%%%%%%%%%%%%%%%%%%%%%%%%%%%%%%%%%%%%
%%%%%%%%%%%%%%%%%%%%%%%%%%%%%%%%%%%%%%%%%%%%%%%%%
%%%%%%%%%%%%%%%%%%%%%%%%%%%%%%%%%%%%

%\bibitem{What?} F. Hammad {\it et al.}, ``What can we learn from the conformal noninvariance of the Klein-Gordon equation?,'' \href{https://www.worldscientific.com/doi/abs/10.1142/S0217751X21502249}{Int. J. Mod. Phys. A {\bf36}, 2150224 (2021)} [\href{https://arxiv.org/abs/2012.12355}{arXiv:2012.12355}].

\bibitem{NeutrinoBook2} M. Sajjad Athar and S.\,K. Singh, \textit{The Physics of Neutrino Interactions} (Cambridge University Press, Cambridge, England, 2020).
%%%%%%%%%%%%%%%%%%%%%%%%%%%%%%%%%%%%%% Synge %%%%%%%%%%%%%%%%%%%%
%%%%%%%%%%%%%%%%%%%%%%%%%%%%%%%%%%%%%%%%%%%%%%%%%%%%%%%%%%
%\bibitem{Synge} J.\,L. Synge, {\it Relativity: The General Theory}, (North-Holland Publishing Company, Amsterdam, 1960).


%%%%%%% Chameleon Field Profile  %%%%%%%
%\bibitem{ChameleonField} T. Nakamura {\it et al}., ``Chameleon Field in a Spherical Shell System,'' \href{https://journals.aps.org/prd/abstract/10.1103/PhysRevD.99.044024}{Phys. Rev. D {\bf99}, 044024 (2019)} [\href{https://arxiv.org/abs/1804.05485}{arXiv:1804.05485}]
%%%%%%%%%%%%%%%%%%%%%%%

%%%%%%%%%%%%%%%%%%%%%%%%%%%%%%%%%%%%%%%%%%%%%%%%%%%%%%%%%%
%%%% Conformal coupling Quintessence models %%%%%%%%%%%%%%
%%%%%%%%%%%%%%%%%%%%%%%%%%%%%%%%%%%%%%%%%%%%%%%%%%%%%%%%%%
%%%%%%%%%%%%%%%%%%%%%%%%%%%%%%%%%%%%%%%%%%%%%%%%%%%%%%%%%%
%%%%%%%%%%%%%%%%%%%%%%%%%%%%%%%%%%%%%%%%%%%%%%%%%%%%%%%%%
%%%%%%%%%%%%%%%%%%%%%%%%%%%%%%%%%%%%%%%%%%%%%%%%%%%%%%%%%%
%%%%%%%%%%%%%%%%%%%%%%%%%%%%%%%%%%%%%%%%%%%%%%%%%%%%%%%%%%
%%%%%%%%%%%%%%%%%%%%%%%%%%%%%%%%%%%%%%%%%%%%%%%%%%%%%%%%%%
%%%%%%%%%%%%%%%%%%%%%%%%%%%%%%%%%%%%%%%%%%%%%%%%%%%%%%%%%%

%\bibitem{MS} F. Hammad, ``Conformal mapping of the Misner-Sharp mass from gravitational collapse,'' \href{https://www.worldscientific.com/doi/abs/10.1142/S0218271816500814}{Int. J. Mod. Phys. D {\bf25}, 1650081 (2016)} [\href{https://arxiv.org/abs/1610.02951}{arXiv:1610.02951}]

%\bibitem{HH} F. Hammad, ``More on the conformal mapping of quasi-local masses: The Hawking-Hayward case,'' \href{https://iopscience.iop.org/article/10.1088/0264-9381/33/23/235016}{Class. Quantum Grav. {\bf33}, 235016 (2016)} [\href{https://arxiv.org/abs/1611.03484}{arXiv:1611.03484}]

%\bibitem{BHWormhole} F. Hammad, ``Revisiting black holes and wormholes under Weyl transformations", \href{https://journals.aps.org/prd/abstract/10.1103/PhysRevD.97.124015}{Phys. Rev. D {\bf97}, 124015 (2018)} [\href{https://arxiv.org/abs/1806.01388}{arXiv:1806.01388}].

 

%\bibitem{BH} F. Hammad, \'E. Mass\'e and P. Labelle, ``Black hole mechanics and thermodynamics in the light of Weyl transformations", \href{https://journals.aps.org/prd/abstract/10.1103/PhysRevD.98.104049}{Phys. Rev. D {\bf98}, 104049 (2018)} [\href{https://arxiv.org/abs/1811.12201}{arXiv:1811.12201}].

%\bibitem{ST} F. Hammad and D. Dijamco, ``More on spacetime thermodynamics in the light of Weyl transformations", \href{https://journals.aps.org/prd/abstract/10.1103/PhysRevD.99.084016}{Phys. Rev. D{\bf99}, 084016 (2019)} [\href{https://arxiv.org/abs/1904.07151}{arXiv:1904.07151}].

%\bibitem{ParallelBH} F. Hammad {\it et al.}, ``Noether charge and black hole entropy in teleparallel gravity,'' \href{https://journals.aps.org/prd/abstract/10.1103/PhysRevD.100.124040}{Phys. Rev. D {\bf100}, 124040 (2019)} [\href{https://arxiv.org/abs/1912.08811}{arXiv:1912.08811}].


%%%%%%%%%%%%%%%%%%%%%%%%%%%%%%%%%%%%%%%%%%%%%%%%%%%%%%%%%%


%%%%%%%%%%%%%%%%%%%%%%%%%%%%%%%%%%%%%%%%%%%%%%%%%%%%%%%%%%
%%%%%%%%%%%%%%%%%%%%%%%%%%%%%%%%%%%%%%%%%%%%%%%%%%%%%%%%%%
%%%%%%%%%%%%%%%%%%%%%%%%%%%%%%%%%%%%%%%%%%%%%%%%%%%%%%%%%%
%%%%%%%%%%%%%%%%%%%%%%%%%%%%%%%%%%%%%%%%%%%%%%%%%%%%%%%%%%

\end{thebibliography}
\end{document}